\documentclass{cls/ws-book9x6}
\usepackage{sty/ws-book-thm}   
\usepackage{sty/ws-book-har}   
\usepackage[pdfpagelabels=false,colorlinks=true,allcolors=black]{hyperref}  

\title{Advances in Quantum Computer Music}

\usepackage{xcolor}
\usepackage{braket}
\usepackage{subcaption}

\makeindex

\begin{document}










\setcounter{page}{1}


\chapter[Developing a Framework for Sonifying Variational Quantum Algorithms]{Developing a Framework for Sonifying Variational Quantum Algorithms: Implications for Music Composition}\label{vqh}

\author{Paulo Vitor Itabora\'i$^{1,2}$, Peter Thomas $^{3}$, Arianna Crippa$^{1,4}$, Karl Jansen$^{1,2}$, Tim Schwägerl$^{1,4}$, María Aguado Yáñez$^{1,5}$ }

\address{\footnotesize
        $^1$Center for Quantum Technologies and Applications -- Deutsches Elektronen-Synchrotron DESY\\
        Platanenallee 6 -- 15738 -- Zeuthen -- Germany\\
        $^2$Computation-based Science and Technology Research Center -- The Cyprus Institute \\
        20 Konstantinou Kavafi Street -- 2121 --Aglantzia -- Cyprus\\
        $^3$Interdisciplinary Centre for Computer Music Research -- University of Plymouth \\
        Drake Circus - PL48AA - Plymouth - United Kingdom\\
        $^4$Instit\"ut für Physik -- Humboldt-Universit\"at zu Berlin \\
        Newtonstr. 15 -- 12489 Berlin --Germany\\
        $^5$Music Technology Group -- Pompeu Fabra University\\
        Roc Boronat, 138 -- 08018 -- Barcelona -- Spain\\
}

\paragraph{Abstract}
     This chapter examines the Variational Quantum Harmonizer, a software tool and musical interface that focuses on the problem of sonification of the minimization steps of Variational Quantum Algorithms (VQA), used for simulating properties of quantum systems and optimization problems assisted by quantum hardware. Particularly, it details the sonification of Quadratic Unconstrained Binary Optimization (QUBO) problems using VQA. A flexible design enables its future applications both as a sonification tool for auditory displays in scientific investigation, and as a hybrid quantum-digital musical instrument for artistic endeavours. In turn, sonification can help researchers understand complex systems better and can serve for the training of quantum physics and quantum computing. The VQH structure, including its software implementation, control mechanisms, and sonification mappings are detailed. Moreover, it guides the design of QUBO cost functions in VQH as a music compositional object. The discussion is extended to the implications of applying quantum-assisted simulation in quantum-computer aided composition and live-coding performances. An artistic output is showcased by the piece \textit{Hexagonal Chambers} (Thomas \& Itabora\'i, 2023).

\section{Introduction}
\label{sec:intro}

The evolution of musical instruments in the 1900's and early 2000's has included electronic and digital technologies in its design and manufacturing, which has greatly impacted how music is composed and played. Modern programming languages such as SuperCollider \cite{SuperCollider}, Pure Data \cite{puckette1996pure}, and Max/MSP \cite{zicarelli1998extensible} have enabled a flexible creation of innovative musical interfaces, allowing performers to dynamically build, modify and interact with new instruments, even while performing on stage \cite{roads1996computer}. This integration of digital and music technologies has introduced new forms of musical expression \cite{nime}. Arguably, a similar effect can occur with quantum technologies and music.


\subsection{Motivation}
\label{subsec:motivation}

The use of sound for representing data is a known strategy for scientific dissemination and the comprehension of time-dependent and high-dimensional datasets. To name some examples, sonification has been used for describing astronimical data, such as the results of the acclaimed experiment on the detection of graviational waves \cite{Zanella2022}. In addition, researchers have used sonification as a tool during the design of multiscale molecular simulations in material physics \cite{Miranda2019SoundedJourney}. Furthermore, sensing instruments have also been traditionally implemented for identifying `invisible' properties of our surroundings through sound, such as detection of ionizing radiation with a Geiger counter \cite{curtiss1950geiger} or more emblematically for music, capacitive coupling of electic fields with a Theremin \cite{glinsky2000theremin}. A more recent example shows that sonification can also be used as real-time auditory feedback in medical applications \cite{corredera2023sonification}. 

Lastly, yet importantly, the well-known impact of sonification in music and sound art should not be overlooked \cite{straebel2010sonification}\cite{xenakis1992formalized}\cite{cage1958experimental}. Integrating innovative technologies in electronic music instruments could create new possibilities for creating and listening to music \cite{meyer1958statistic}. As stated by the instrument inventor Anton Walter Smetak: \textit{``A new world requires new people and new music - and to that end, different musical instruments"} \footnote{Quote translated from Portuguese: ``Um novo mundo requer homens novos e uma musica nova - e para isso, instrumentos musicais diferentes" \cite[p.xi]{Scarassatti2001}}

With the emerging field of quantum computing, a novel technology could be integrated to offer the possibility of going in different directions of musical expression at many levels of the artistic process.

In conclusion, creating auditory displays for comprehending, disseminating, and assisting research in quantum technologies is a problem that should be explored in depth by Quantum Computer Music academics and eventually the industry.

In another note, circuit-based quantum computing with qubits deals with the control and interaction of two-level quantum systems, usually coupled together to achieve complex behavior and physical phenomena, which could be difficult for classical machines to simulate. Likewise, a similar analogy could be made when considering acoustic musical instruments. According to Fletcher and Rossing's definition:

\textit{“In most instruments, sound production depends upon the collective
behavior of several vibrators, which may be weakly or strongly coupled
together. This coupling, along with nonlinear feedback, may cause the instrument as a whole to behave as a complex vibrating system, even though the individual elements are relatively simple vibrators.” }\cite{fletcher2012physics}

This analogy provides a motivation layer for the interpretation of qubits as individual elements, which are coupled together in order to compose a complex structure that could be used to generate organized sound \cite{OrganisedSound}.

\subsection{Previous work}
\label{subsec:previous_work}
\subsubsection{Interfacing Quantum Computing with Musical Interfaces}
\label{subsubsec:interfacingQC}
There are a relatively small number of early studies on the topic of the sonification of quantum algorithms, either for educational or artistic purposes. In general, the literature on Quantum Computer Music is very recent. At the time of writing, the most widely adopted references can be found in two books \cite{mirandaQuantumComputerMusic2022}\cite{mirandaQuantumComputingArts2022}, and conference proceedings \cite{isqcmc}, among other works of the same period.
By surveying the works related to applications of sonification and their software implementations, it is possible to verify how researchers have addressed the problem of integrating quantum computing frameworks with music programming. The approaches can be divided into two main categories: Music-Oriented and Quantum-Oriented.

\paragraph{A Music-Oriented Approach} 
In frameworks centered in music programming languages, a common challenge lies in the need of a platform for a) generating quantum circuits and b) running simulations or connecting to quantum hardware. In the literature, it is possible to find implementations of microqiskit in Max/MSP \cite{Hamido2022} and statevector simulators in Ableton Live \cite{Weaver2022}. Additionally, there is a software application that listens to quantum circuit instructions sent via Open Sound Control (OSC) Protocol \cite{OSC} for running circuits in quantum systems \cite{Hamido2023}. The last mentioned application has been used to drive the quantum backend for the sonification of a Bloch sphere, implemented in a web-based audio application \cite{q1synth}.
Due to the current limitations of the implementations of these simulators and qasm code generators, the applications are currently restricted to relatively simple quantum circuits and manipulations.

\paragraph{A Quantum-Oriented Approach} 
Conversely, it is more common to find Quantum-Centered applications. This means that the main implementation is (typically) realised in Python, and then the challenge changes to finding separate modules and software frameworks to generate sound and music reliably. Researchers have utilized this approach to use more complex quantum circuits in sonification problems and music composition. To list some examples from the literature, quantum algorithms have been used to generate or trigger MIDI notes \cite{MirandaHari2023}\cite{ISQCMC_yanez}, musical scores \cite{Miranda2022TeachingQT}, and even store and retrieve audio files \cite{Itaborai2022}\cite{itaborai2023towards}. Particularly, the latter reference uses Python as a SuperCollider client \cite{pythonsupercollider}, by implementing formatted OSC messages to instantiate synthesizers and control the audio engine. This is the approach chosen for the implementation presented in this work.

\subsubsection{Sonifying the Ising Model}
\label{subsubsec:sonifying_ising_model}
Considering the previous work published as a chapter of the preceding Quantum Computer Music book \cite{Clemente2022}, in Section 3, entitled \textit{``The sound of the Ising Model"}, the authors utilize a Variational Quantum Algorithm (VQA) such as the one described in section \ref{par:vqe} to extract observables\footnote{i.e., any physical quantity that can be measured, such as energy, momentum, spin, etc.} from the Ising Model and create a simple sonification mapping, where the values represent frequencies of a sound spectrum mapped to a perceptual listening range.

They propose an additive synthesis approach for sonifying the Ising Model using two different control signals. The first method maps the physical properties of the Ising Model - Energy eigenvalues and Magnetizations - as frequencies and amplitudes, respectively. Furthermore, the spin coupling \textit{h} is used as a time variable, that evolves the system through a phase transition. In other words, in their example, for each timestamp \(t_n\) there were 16 eigenvalues \(E_{n}\) and 16 eigenstates \(\ket{\psi_{n}}\) (from where the magnetizations \(M_{n}\) were computed) from a VQA run with a specified coupling \(h(n)\).

More compellingly (for the purposes of this chapter), in the map proposed in Sec. 3.1.2 (\textit{``Use the callback results"}), Instead of using only the output of the algorithm, where in principle the observables could have been computed by a classical algorithm they \textit{``unfold"} the time \(t_n\) and adjust their lenses to the intermediary steps of the minimization process, and compute the eigenvalues \(E_n\) at every iteration. Hence, they infer the sonification of the VQA itself.

In sequence of \cite{Clemente2022}, \cite{ISQCMC_itaborai} expands on the suggestion proposed above, and includes the sonification of the quantum states \(\ket{\psi_n}\) at intermediate stages of a VQA optimization, which becomes a primary point of interest on the sonification map. By prototyping a system that realizes the proposed sonification for a Variational Quantum Eigensolver (VQE) algorithm and produces audible output, the Variational Quantum Harmonizer (VQH) was conceived.

In \cite{ISQCMC_itaborai}, there is a focus on introducing the topic, obtaining musical intuition, and also a discussion on an artistic output in the form of a music study, named \textit{``Dependent Origination"}.

The aim of this work is to provide an overview of the second version of the VQH implementation, as well as broaden the concepts introduced, with the intention of allowing the reader to incorporate and use the VQH implementation in their own projects. Furthermore, there is a brief discussion on the use of VQH for creating music compositions as well as implications and intricacies of the artistic process for creating the music piece \textit{``Hexagonal Chambers"}(See Section \ref{subsec:hexagonal_chambers}).

\section{Variational Quantum Harmonizer}
\label{sec:vqh_overview}

The \textit{Variational Quantum Harmonizer} is a hybrid quantum-classical sonification-based musical instrument, implemented in Python and using Qiskit \cite{Qiskit} for running quantum algorithms and generating data that is sonified using Supercollider \cite{SuperCollider} as its main sound engine system. In other words, the VQH is a sonification tool for mapping Variational Quantum Algorithms into sound.

\paragraph{Variational Quantum Eigensolver}
\label{par:vqe}

\begin{figure}[ht!]
    \centering
    \includegraphics[width=1\textwidth]{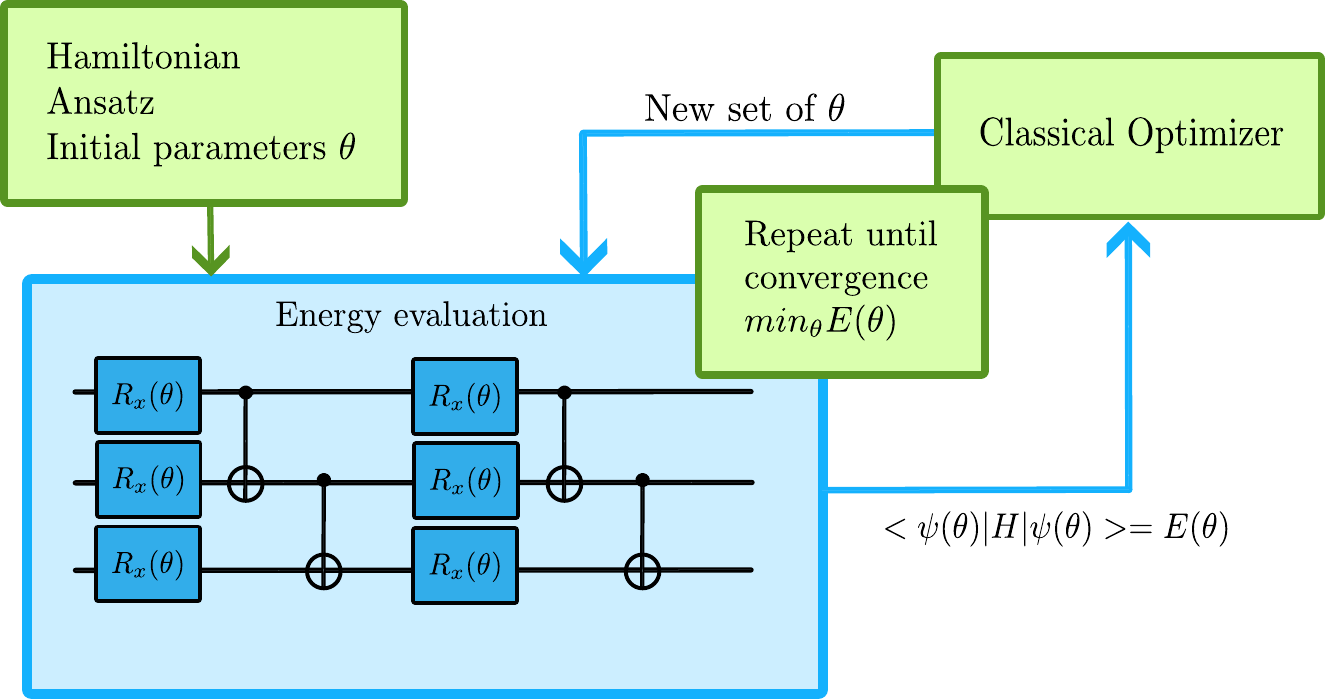}
    \caption{Schematic illustration of the Variational Quantum Eigensolver (VQE) algorithm.}
    \label{fig:vqealg}
\end{figure}

The Variational Quantum Eigensolver (VQE)\cite{Peruzzo2014}, (Fig.\ref{fig:vqealg}) is a hybrid algorithm for simulations and optimisation problems, that integrates classical and quantum computing paradigms to determine the ground state and some excited states of a given physical system. The algorithm is based on the variational method of quantum mechanics, i.e. finding the lowest energy state via an approximation strategy.
In order to apply this technique, a quantum circuit is defined with a set of parametrized and entangling gates and with an initial guess for the parameters $\theta$.
By exploiting this initial configuration, the quantum processor calculates the expectation value of an observable of the system, typically the Hamiltonian, getting the value of the cost function $E(\theta)=\bra{\psi(\theta)}H\ket{\psi(\theta)}$. Then, a classical optimizer iteratively refines the initial parameters, with a new set of $\theta$. This procedure is repeated until the optimization reaches convergence and a minimum of the cost function is found\footnote{Notice, however, that this algorithm can also get trapped in local minima or constant plateau landscapes}.

In essential terms, the VQH has two distinct levels of approach to its control interface and subsequent workflows. In a higher-level routine, the user can load datasets containing formatted results of a VQE experiment and map the data into various sonification mappings to listen to the results. Alternatively, the user is encouraged to design their own experiments from the ground up, using the interface, with the goal of obtaining sonic intuition on the way a problem of interest works, and most importantly how the VQE is performing to achieve that result. This represents the broader methodology of the VQH application as a whole, towards a system that can be used both to assist in the analysis and interpretation of quantum simulations or to compose music.

\subsection{The VQH prototype and overview}
\label{subsec:vqh_prototype}

For the first part of this project, an implementation that realizes the proposed pipeline (a real-time sonification tool for VQE experiments) was reached for a simple problem and also with a direct encoding as a quantum circuit and a clear mapping into sound. 

In more detail, the user controls and designs parameters of a \(n \times n\) square matrix that defines a Quadratic Unconstrained Binary Optimization (QUBO) problem \footnote{In practice, since the QUBO matrix is symmetric, only the upper triangle elements need to be defined}.

\paragraph{QUBO}
Quadratic unconstrained binary optimization (QUBO)\cite{date2021qubo} is a combinatorial optimization problem, i.e. the problem of finding the best solution from all feasible solutions. It is characterised by a cost function (Eq.\ref{eq:qubo}), comprising both linear and quadratic terms. Each variable $n_i$ uniquely contributes to the system with its linear coefficients $a_i$, while interacting with the other agents through the quadratic coefficients $b_{ij}$. Effectively solving this problem means identifying (or approximating) a configuration that minimises the cost function.
\begin{equation}\label{eq:qubo}
    Q(a,b,n)=\sum_i^N a_i n_i + \sum_{i,j} b_{ij} n_i n_j \ \ \ \text{with}\; n \in \{0,1\}^N\;.
\end{equation}

\vspace{5pt}

Then, the problem is transformed into an Ising Hamiltonian `\(H\)', by mapping the binary variables \(n_i\) to Pauli \(Z\) operators (Eqs. \ref{eq:agent_to_spin}-\ref{eq:qubotoqubit}). This characterizes an energy landscape from where the ground state can be approached by means of a VQE-based optimization.
Subsequently, at each iteration of the optimization process, the current configuration is extracted and sampled by a callback function to obtain a probability distribution that describes the current state of the system, as well as an intermediary-stage computation of the energy expectation value for the designed Hamiltonian.

\begin{align}
    n_i &\Rightarrow \sigma^Z_i\ket{n_i} = (1-2n_i) \ket{n_i\label{eq:agent_to_spin}}\;,\\
    b_{ij} &\Rightarrow \beta_{ij} = b_{ij} \label{eq:b_to_beta}\;,\\ 
    a_i &\Rightarrow \alpha_i \;= -2a_i - \sum_{j\ne i}b_{ij}
    \label{eq:a_to_alpha}\;,
\end{align}

\begin{align}
H(\alpha, \beta, \sigma)=\sum_i^N \alpha_i \sigma^Z_i + \sum_{i,j} \beta_{ij}\sigma^Z_i \sigma^Z_j 
\label{eq:qubotoqubit}\;.
\end{align}

\begin{figure}[ht!]
    \centering
    \includegraphics[width=.8\textwidth]{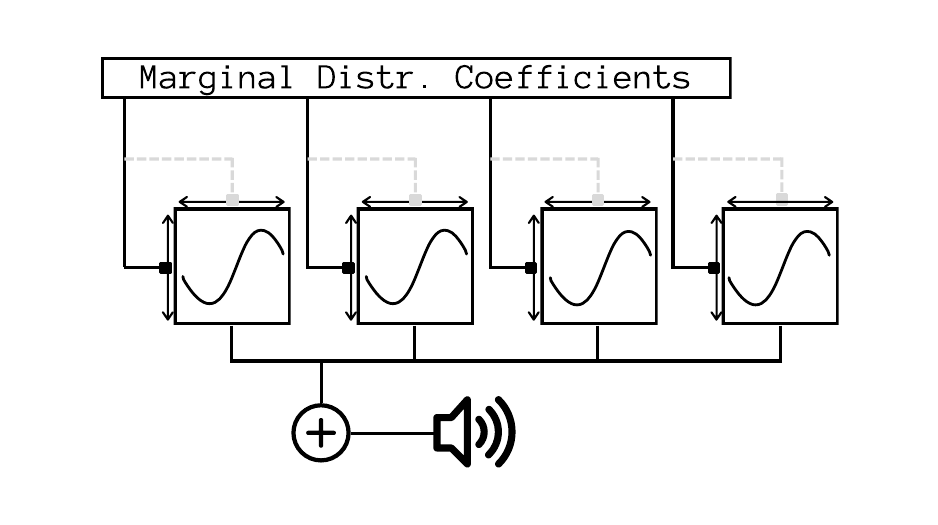}
    \caption{Additive Synthesis Mapping}
    \label{fig:additive}
\end{figure}

Having obtained a sample from the quantum state's probability distribution and corresponding expectation value, the first stage of sonification mapping is applied, where a decoding procedure - which will be referred to in this text as \textit{`Basis Protocol'} - is applied to represent a qubit sequence as a group of \(n\) sounding ($\ket{1}$) or non-sounding ($\ket{0}$) notes (see Section \ref{subsubsec:harp_analogy}). This protocol is carried out by obtaining the \textit{marginal distribution} of each individual qubit, aggregating the amplitudes of the basis states that contribute for each respective note to play. As a result, a set of \(n\) relative amplitudes is achieved.
By collecting these sets over many iterations, it is possible to build \(n\) streams of data to control, for instance, the amplitude of \(n\) sinusoidal oscillators with distinct frequencies (see Sec. \ref{fig:additive}).

\paragraph{Marginal Distribution and the Basis Protocol} This protocol defines the main sonification strategy used in this text. To establish a comparative relation: in data visualization, the content to be seen needs to be grouped and arranged into a mesh, geometry, or grid before a data attribute is mapped into color and rendered as an image. In the Basis Protocol, the `sonic geometry' is the qubit marginal distribution of the sampled state. The marginalization process discards the information related to conditional probability in joint distributions, as the one of a multi-qubit system. In other words, it is the probability of a qubit being in a state \(\ket{1}\) or \(\ket{0}\), independently of the states of other qubits. To illustrate, consider the 3-qubit example below (Eqs. \ref{eq:marginal_state}-\ref{eq:marginal_q2}).

\begin{align}
    \ket{\psi_{abc}} = \sqrt{\frac{3}{4}}\ket{100} + &\sqrt{\frac{3}{16}}\ket{011} + \sqrt{\frac{1}{16}}\ket{101} \label{eq:marginal_state}\;,\\
    P_{marg}(q_a = \ket{1}&) = 3/4 + 1/16 = 13/16 \label{eq:marginal_q0}\;,\\
    P_{marg}(q_b = \ket{1}&) = 3/16\label{eq:marginal_q1}\;,\\
    P_{marg}(q_c = \ket{1}&) = 3/16 + 1/16 = 1/4\label{eq:marginal_q2}\;.
\end{align}

\subsubsection{Artistic Intuition: The Harp analogy}
\label{subsubsec:harp_analogy}

To obtain musical intuition on the Basis Protocol, we introduce an analogy, using a figurative Harp as an object where the qubits are represented as strings. Consider a simple 3-qubit example.

\begin{figure}[h!]
    \centerline{
    \includegraphics[width=0.3\textwidth]{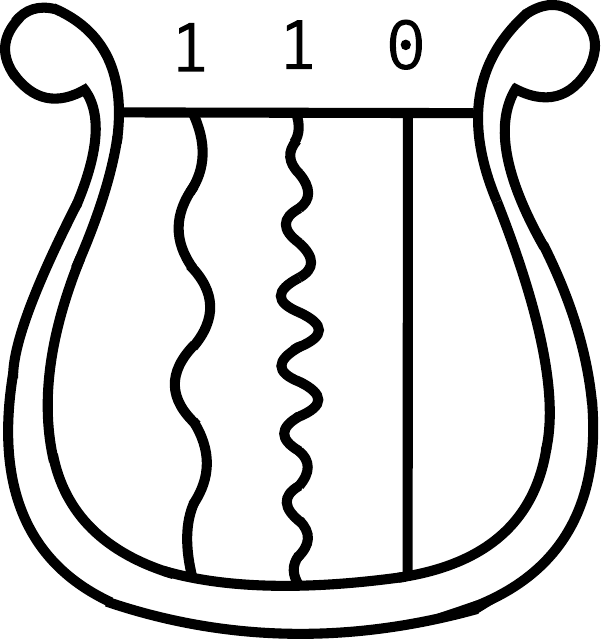}}
    \caption{A 3-qubit harp}
    \label{fig:qubo_harp}
\end{figure}

In this 3-qubit Harp, the two first qubits are playing, while the third one remains silent. To configure this scenario as the ground state following the execution of the VQE algorithm, a cost function has to be designed with a QUBO coefficient matrix, so that it yields the desired outcome. There are two strategies for formulating a QUBO problem, both capable of achieving this end result. The first one involves tuning the linear coefficients $a_i$, assigning a value of $-1$ to the playing qubits and $1$ to the silent one. This means that the third qubit's linear coefficient is being \textit{penalized}, since it increases the cost function, whereas the algorithm aims at minimizing it. This first QUBO matrix would be then implemented as follows in Eq.\ref{eq:qubotoyexample1} \footnote{Note that the variables $n$ can take either value $1$ or $0$, so that $n_i^2 = n_i$.},

\begin{equation}
    Q(n) = - n_1 - n_2 + n_3 = \begin{pmatrix}
n_1 & n_2 & n_3
\end{pmatrix} \begin{pmatrix}
      -1 &\:\:\: 0 &\:\:\:  0\; \\
\:\:\: 0 &      -1 &\:\:\:  0\; \\
\:\:\: 0 &\:\:\: 0 &\:\:\:  1\;
\end{pmatrix} \begin{pmatrix}
n_1 \\ n_2 \\ n_3
\end{pmatrix}
    \label{eq:qubotoyexample1}.
\end{equation}

The second approach builds the QUBO problem tuning the quadratic coefficients $b_{ij}$ instead. By penalizing the interaction between the silent qubit with respect to the playing ones, and favouring the interaction between the first two, as in Eq.\ref{eq:qubotoyexample2}, the desired ground state can be reached.

\begin{equation}
    Q(n) = \begin{pmatrix}
n_1 & n_2 & n_3
\end{pmatrix} \begin{pmatrix}
\:\:\: 0 &      -1 &\:\:\: 1\; \\
      -1 &\:\:\: 0 &\:\:\: 1\; \\
\:\:\: 1 &\:\:\: 1 &\:\:\: 0\;
\end{pmatrix} \begin{pmatrix}
n_1 \\ n_2 \\ n_3
\end{pmatrix}.
    \label{eq:qubotoyexample2}
\end{equation}

Another way of visualizing the quadratic coefficients is by looking into the Ising Hamiltonian that is created from the QUBO coefficients (Eqs. \ref{eq:a_to_alpha}-\ref{eq:qubotoqubit}). In this simple example, the Ising Model would characterize 3 spins. Translating the Harp analogy to the Ising Model, a silent note could be represented by a spin-down state (\(\ket{\downarrow}\)), whereas a playing note is encoded as a spin-up (\(\ket{\uparrow}\)). This allows for a physical interpretation of the quadractic coefficients, where the coupling determines the disposition of the spin pair with respect to each other. From equation \ref{eq:qubotoqubit}, a negative \(\beta\) value would favor a ferromagnetic alignment ($\uparrow\uparrow$, $\downarrow\downarrow$), while a positive value benefits anti-ferromagnetic configurations ($\uparrow\downarrow$, $\downarrow\uparrow$)\footnote{{Hence, one should be careful when designing with quadratic coefficients, since the alignment determines the direction of each spin, but not its orientation in respect to an external field. Sometimes, there is a need to introduce some linear correction factors, to ensure a desirable orientation and thus ground state. See Sec. \ref{subsubsec:breaking_deg}.}}.

\begin{figure}[h]
    \centerline{
    \includegraphics[width=0.3\textwidth]{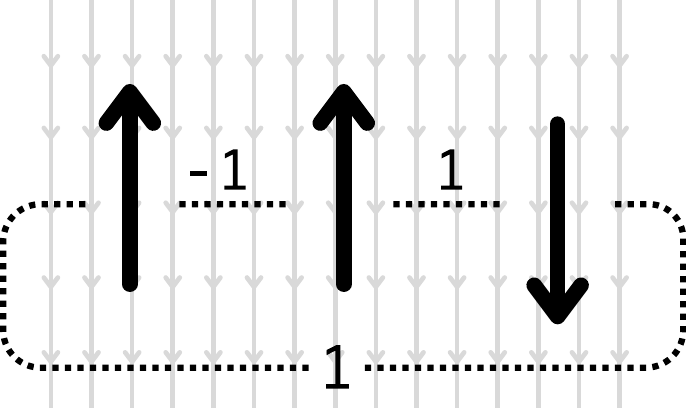}}
    \caption{A 3-spin Ising harp}
    \label{fig:ising_harp}
\end{figure}

\section{Designing Chords}
\label{sec:designing_chords}

Expanding from the analogy above, the first subsequent idea is to explore simple musical scales (e.g., Pentatonic, Pythagorean, Modal, Diatonic, etc..), and attempt to design cost functions in which the ground state is known, enabling us to encode, for example, a chord.

\subsection{Cmajor Chord}
\label{subsec:cmajor}

As a starting point, let us create a 12-tone Chromatic Scale. 

\begin{figure}[ht]
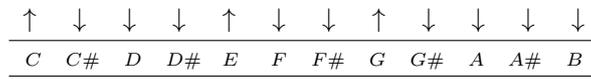

    \centering
    \begin{tabular}{cccccccccccc}

        $\uparrow$\, &  $\downarrow$\, & $\downarrow$\, & $\downarrow$\, & $\uparrow$\, & $\downarrow$\, & $\downarrow$\, & $\uparrow$\, & $\downarrow$\, & $\downarrow$\, & $\downarrow$\, & $\downarrow$
    \\

    \end{tabular}
    
    \begin{tabular}{cccccccccccc}
    \hline
     \scriptsize{$C$}\!  & \!\scriptsize{$C\#$}\! & \!\scriptsize{$D$}\! & \!\scriptsize{$D\#$}\!\! & \!\scriptsize{$E$} & \scriptsize{$F$}\! & 
     \!\scriptsize{$F\#$}\! & \!\scriptsize{$G$}\! & \!\scriptsize{$G\#$}\! & \!\scriptsize{$A$}\! & \!\scriptsize{$A\#$}\! & \!\scriptsize{$B$}
     \\\hline
    \end{tabular}
    \caption{Spin configuration of a C Major Chord}
    \label{fig:isingharp}
\end{figure}

Accordingly, 12 qubits are assigned as shown in fig \ref{fig:isingharp}. Then, the following cost function, shown in Eq. \ref{eq:qubocmaj}, is designed to be minimized into a \textit{C Major} Chord using only linear coefficients for now.

\begin{equation}
    Q(n_{Cmaj}) = - n_C - n_E-n_G+\sum_{k\notin Cmaj} n_k
    \label{eq:qubocmaj}\quad\!.
\end{equation}

When \ref{eq:qubocmaj} is translated into a qubit-based Hamiltonian, the following set of Pauli operators is achieved (Eq. \ref{eq:cmaj_paulis}\footnote{To keep the same convention from the harp analogy, the qubits are being ordered in Big-endian (left-to-right) form in equation \ref{eq:cmaj_paulis}, contrary to Qiskit's Little-endian notation.}$^,$\footnote{Also note that the signs are swapped compared to the QUBO coefficients of Eq. \ref{eq:qubocmaj}, as \(Z\ket{\downarrow} = \ket{\downarrow}\) and \(Z\ket{\uparrow} = -\ket{\uparrow}\)\ (Eq. \ref{eq:agent_to_spin}). This ensures that \(\ket{\uparrow\downarrow\downarrow\downarrow\uparrow\downarrow\downarrow\uparrow\downarrow\downarrow\downarrow\downarrow}\) is the ground state for Eq. \ref{eq:cmaj_paulis}}).
\setlength{\jot}{-3pt}
\begin{equation}
    \begin{aligned}
        2\big[ \;& -\  Z  I I I I I I I I I I I\\
        & +\  I Z I I I I I I I I I I\\
        & +\  I I Z I I I I I I I I I\\
        & +\  I I I Z I I I I I I I I\\
        & -\  I I I I Z I I I I I I I\\
        & +\  I I I I I Z I I I I I I\\
        & +\  I I I I I I Z I I I I I\\
        & -\  I I I I I I I Z I I I I\\
        & +\  I I I I I I I I Z I I I\\
        & +\  I I I I I I I I I Z I I\\
        & +\  I I I I I I I I I I Z I\\
        & +\  I I I I I I I I I I I Z\ \big]\\
    \end{aligned}
    \label{eq:cmaj_paulis}
\end{equation}

\setlength{\jot}{3pt}

We use the COBYLA \cite{powell1994direct} optimizer and Qiskit's EfficientSU2 ansatz to perform a VQE minimization on this energy profile.

\begin{figure}[ht!]
    \centering
    \includegraphics[width=.8\textwidth]{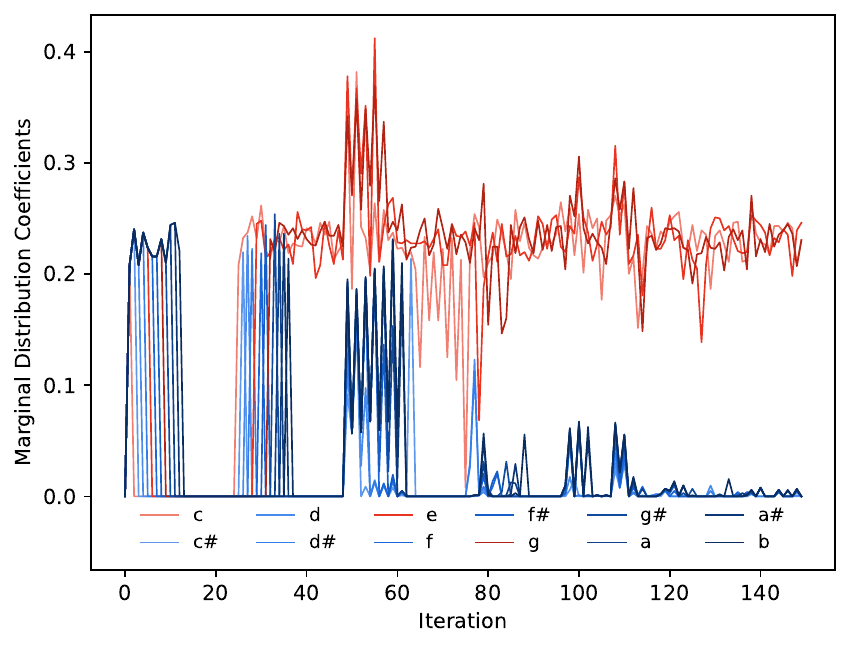}
    \caption{Marginal Distribution Evolution for \(C_{maj}\)}
    \label{fig:cmajorlinear}
\end{figure}

As mentioned in the section above, at each iteration, the current state is sampled. To monitor which notes are playing at that stage, the marginal distribution of the sample is computed. The progression of this marginal distribution across 150 iterations of the optimization is depicted in Fig. \ref{fig:cmajorlinear}.

\begin{figure}[h]
    \centerline{\includegraphics[width=0.9\linewidth]{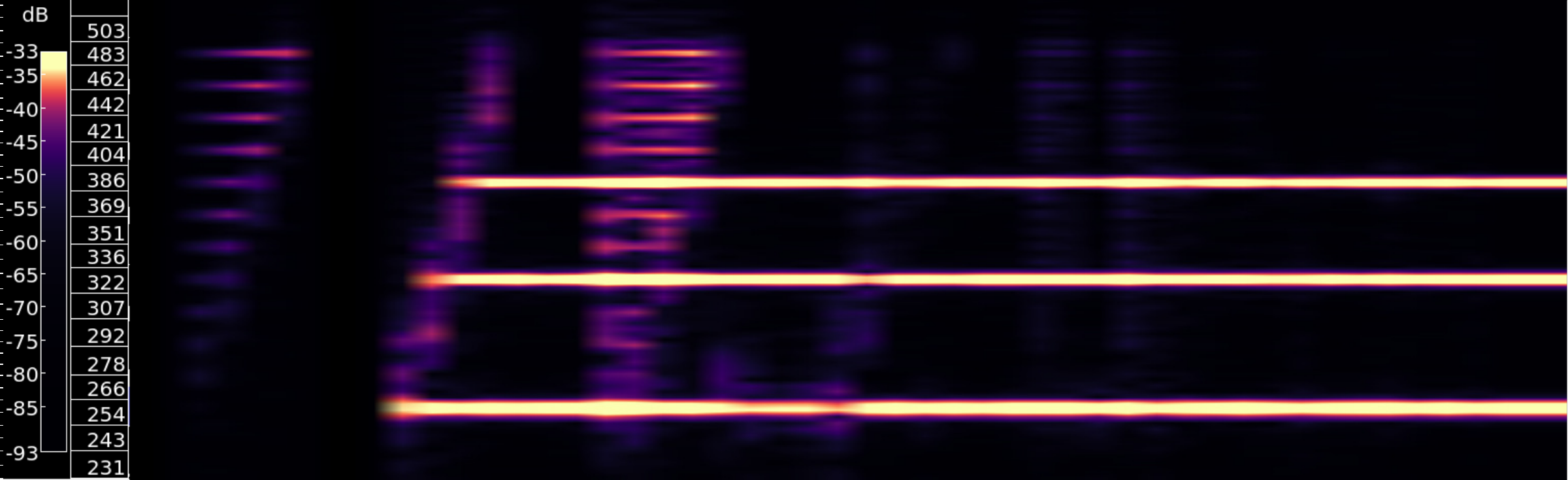}}
    \caption{Spectrogram of \(C_{maj}\) sonified with additive synthesis}
    \label{fig:cmajorlinear_spec}
\end{figure}

The coefficients are then mapped into an additive synthesis with 12 oscillators (tuned according to the chromatic scale), and their respective amplitudes are controlled over time. A spectrogram of the resulting sound is shown in Figure \ref{fig:cmajorlinear_spec}

One can see how the optimizer managed the minimization process and steered the configuration space toward the anticipated solution - a sounding C major.

\subsection{Chord Progressions}
\label{subsec:chord_progressions}

In the example above, the experiment was designed in a way that it starts in ``silence", meaning the optimization begins with all qubits in the \(\ket{0}\) state. However, it is technically possible to choose any other valid state as the initial condition for the VQE, \textit{including one that encodes a different note or chord}. 
Consequently, a problem could be designed in a way that an arbitrary initial chord $C_1$ transitions to a new chord $C_2$.

\begin{figure}[h]
    \centerline{\includegraphics[width=0.6\linewidth]{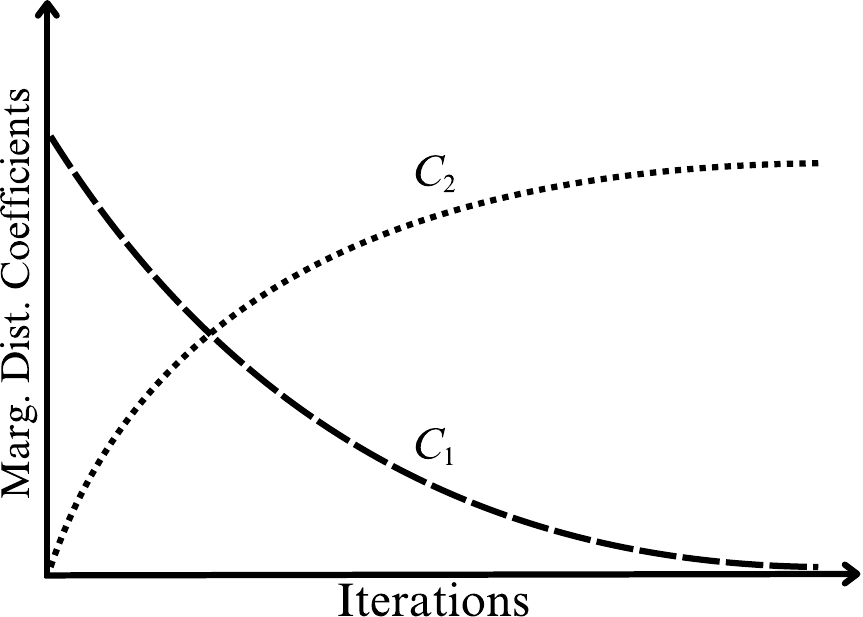}}
    \caption{Changing the initial condition}
    \label{fig:chord_transition}
\end{figure}

\begin{figure}[h]
    \centerline{\includegraphics[width=0.7\linewidth]{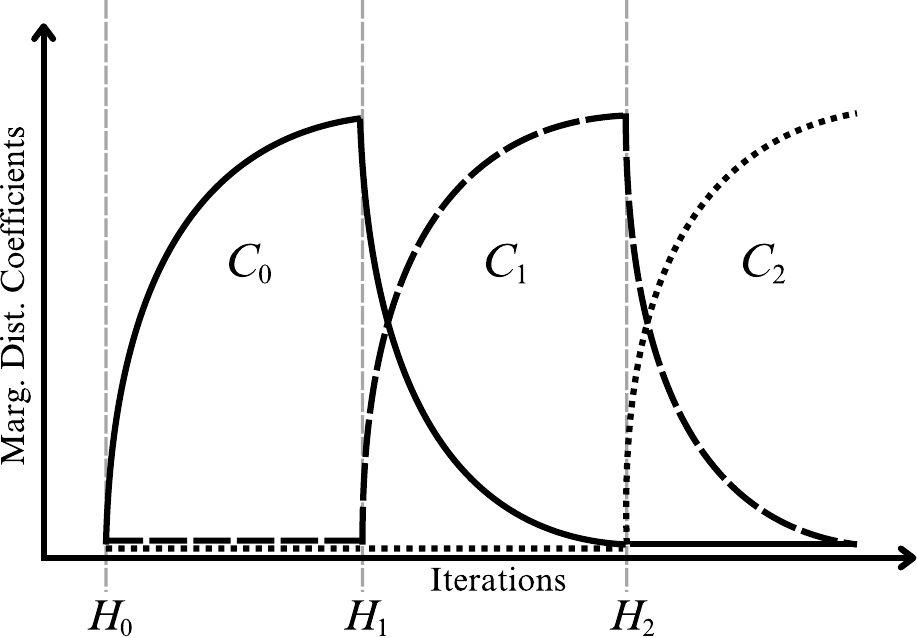}}
    \caption{Creating chord progressions}
    \label{fig:chord_progression}
\end{figure}

Expanding this idea, the next step would be to concatenate different VQE experiments by using the results of a previous run as the initial conditions for a new one, thus creating long chord progressions, timbral transitions, or more complex sonic transformations. These would then be controlled by the changes in the energy profile of the system, characterized by a time-dependent Hamiltonian (Fig. \ref{fig:chord_progression}).

\subsubsection{Example: I-IV-V-I Progression}
Consider the following cost functions \(Q_{1-4}\) below (Eqs. \ref{eq:q1}-\ref{eq:q4}).

\begin{align}
    Q_1(n^I) = - n_C - n_E-n_G+\sum_{k\notin Cmaj} n_k \label{eq:q1}\;,\\
    Q_2(n^{IV}) = - n_F - n_A-n_C+\sum_{k\notin Fmaj} n_k\;,\\
    Q_3(n^V) = - n_G - n_B-n_D+\sum_{k\notin Gmaj} n_k\;,\\
    Q_4(n^I) = Q_1(n^I) \label{eq:q4}\;.
\end{align}

\begin{figure}[ht!]
    \centering
    \includegraphics[width=.7\textwidth]{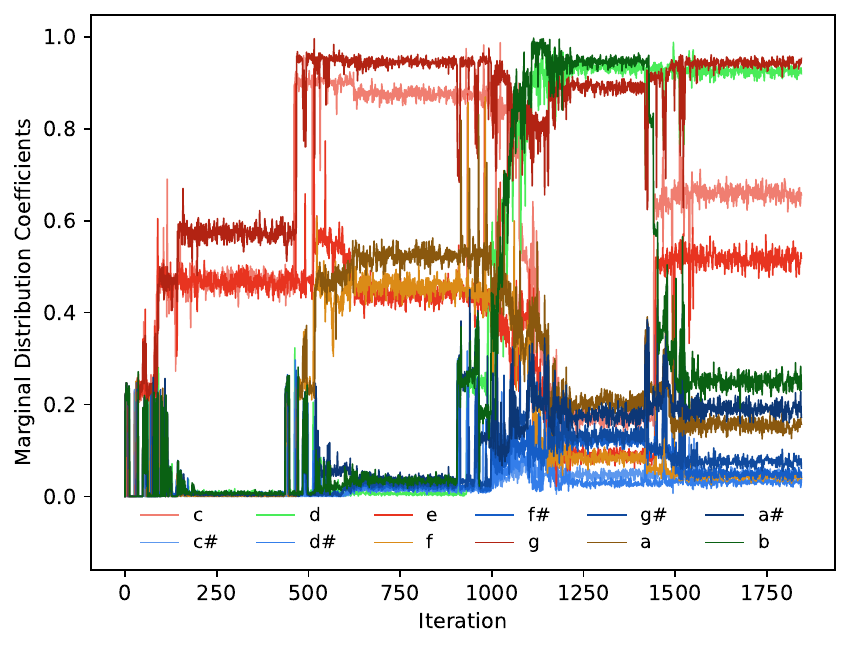}
    \caption{Marginal Distribution Evolution of a I-IV-V-I Progression}
    \label{fig:iivvilinear}
\end{figure}

Then, 4 experiments are prepared with 512 iterations, and identical setup (COBYLA, EfficientSU2). However, the initial point of \(Q_{n+1}\) is the solution of \(Q_n\). By concatenating the iterations of all runs, we are left with a classic I-IV-V-I chord progression \ref{fig:iivvilinear}.

\begin{figure}[h]
    \centerline{\includegraphics[width=0.9\linewidth]{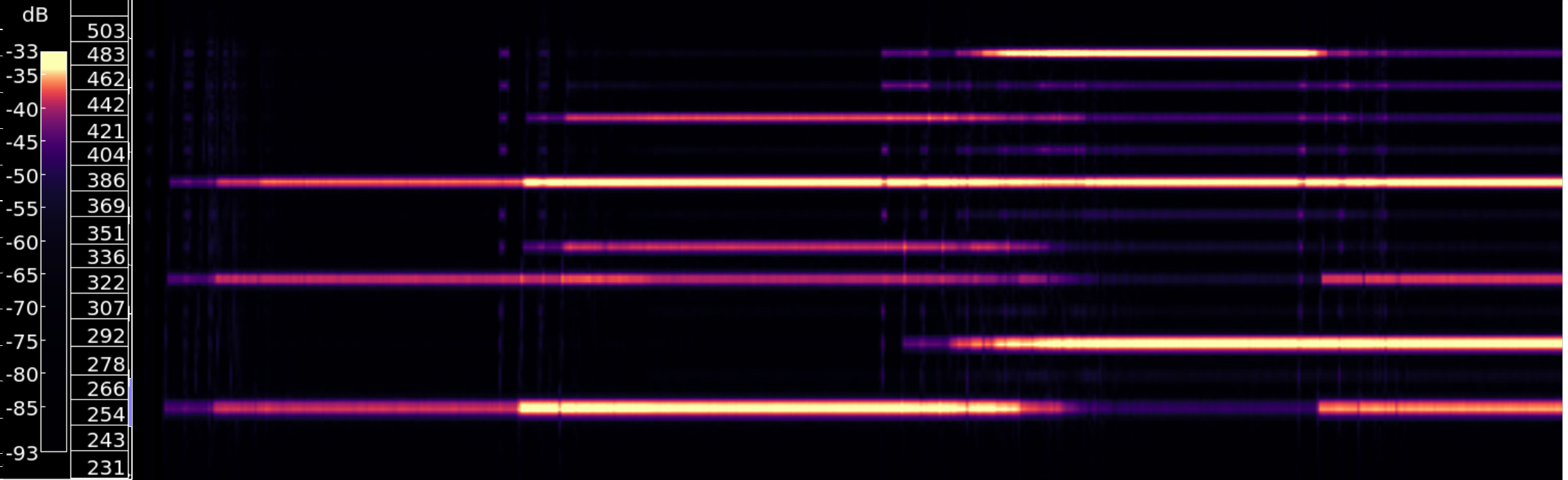}}
    \caption{Spectrogram of I-IV-V-I progression with additive synthesis}
    \label{fig:iivvilinearspec}
\end{figure}

It is possible to note that, by changing the initial point from ``silence" to a different state, the minimization process will necessarily have another path, which might produce the appearance of ``intrusive" notes (such as the A\# in the last chord), or the persistence of previous notes where a decrease in the note's amplitude may not provide significant changes in the minimization task of the current chord.

\subsubsection{Adiabatic Progressions}
\label{subsubsec:adiabatic_progression}

As an additional option, it is possible to consider smoothing down the chord transition. For instance, one could consider a Hamiltonian that is a combination of one chord and another, and use this Hamiltonian as halfway point. Ultimately, with enough intermediary steps, the aim would be to slow the progression of the chord until it evolves \textit{adiabatically}. In fact, this is a technique used in the literature to attempt to improve the VQE minimization, called \textit{Adiabatically Navigated VQE} \cite{Matsuura_2020}. In other words, we design a time-dependent Hamiltonian that guides the solution from \(H_{initial}\) to \(H_{final}\) (Eq. \ref{eq:adiabatic}). 

\begin{equation}
    H(t) = (1-t)H_{initial} + tH_{final}
    \label{eq:adiabatic}\;.
\end{equation}

\begin{figure}[ht!]
    \centerline{\includegraphics[width=.7\textwidth]{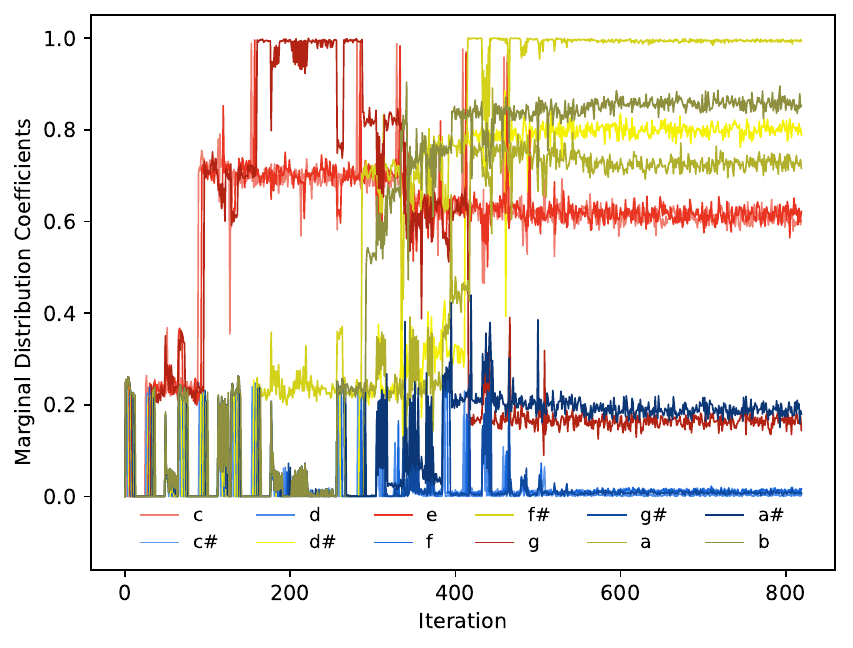}}
    \caption{Adiabatic Chord Progression from $\ket{C_{maj}}$ to $\ket{B7/D\#}$}
    \label{fig:adiabatic}
\end{figure}

For example, one could start with a $\ket{C_{maj}}$ configuration and make an adiabatic transition to $\ket{B7/D\#}$ (Fig. \ref{fig:adiabatic})

\subsection{Degenerate Solutions}
\label{subsec:degenerate_sol}

In sequence, let us analyze a different cost function (Eqs. \ref{eq:ising_cmaj_degenerate}-\ref{eq:ising_linear_correction}), where some quadratic terms are used. The terms, when translated into an Ising problem, indicate a coupling between nearest-neighbor spins in a linear lattice, using periodic boundary conditions.

\begin{equation}
        b_{kl} =
\begin{cases}
    1, &\begin{aligned}
        & k \in Chord;\, l \notin Chord \\ & k \notin Chord;\, l \in Chord
    \end{aligned} \\
    -1, & \text{otherwise}
\end{cases}
    \label{eq:ising_cmaj_degenerate}
\end{equation}

\begin{equation}
        a_{k} = -\frac{1}{2}\sum_{k\neq l}b_{kl}
        \label{eq:ising_linear_correction}
\end{equation}

\begin{figure}[h]
    \centerline{\includegraphics[width=\linewidth]{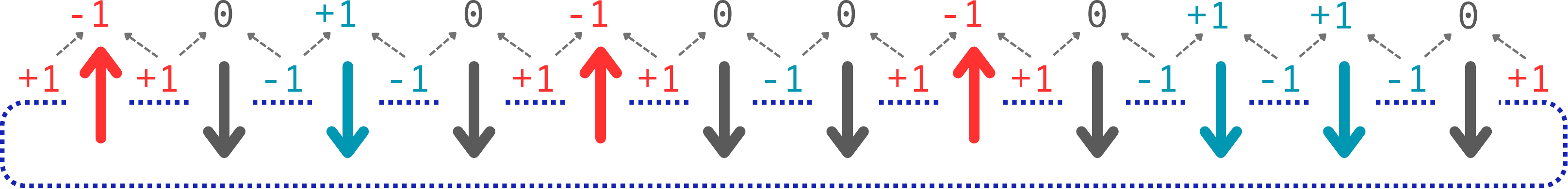}}
    \caption{Spin configuration of a \(C_{maj}\) with nearest-neighbour coupling}
    \label{fig:ising_cmajor_degenerate}
\end{figure}

It is possible to verify that this cost function, together with the linear coefficients \(a_k\), are designed so that the solution has two equally good configurations that minimize it: $\ket{\uparrow\downarrow\downarrow\downarrow\uparrow\downarrow\downarrow\uparrow\downarrow\downarrow\downarrow\downarrow}$ and $\ket{\downarrow\uparrow\uparrow\uparrow\downarrow\uparrow\uparrow\downarrow\uparrow\uparrow\uparrow\uparrow}$ (Eq. \ref{eq:cmaj_paulis_quadratic}). In result, every time the experiment is run, the algorithm can reach a state where \textit{Cmaj} is playing, or where all notes are playing except the \textit{Cmaj} constituting notes are playing. The resulting state can also be influenced by the choice of the classical optimizer, initial conditions and ansaetze. To keep the same terminology as from the previous work, this state could be named \textit{"anti-Cmaj"}.

\setlength{\jot}{-3pt}
\begin{equation}
    \begin{aligned}
        & \qquad\qquad\qquad\qquad\qquad\quad\; + Z Z I I I I I I I I I I\\
        & \qquad\qquad\qquad\qquad\qquad\quad\;- I Z Z I I I I I I I I I\\
        & \qquad\qquad\qquad\qquad\qquad\quad\;-  I I Z Z I I I I I I I I\\
        & -\  Z I I I I I I I I I I I   \quad \quad\; \quad+  I I I Z Z I I I I I I I\\
        & +\  I I Z I I I I I I I I I   \quad \quad\; \quad+  I I I I Z Z I I I I I I\\
        & -\  I I I I Z I I I I I I I   \quad \quad\; \quad-  I I I I I Z Z I I I I I\\
        & -\  I I I I I I I Z I I I I   \quad +\; \quad+\: I I I I I I Z Z I I I I\\
        & +\  I I I I I I I I I Z I I   \quad \quad\; \quad+  I I I I I I I Z Z I I I\\
        & +\  I I I I I I I I I I Z I   \quad \quad\; \quad-  I I I I I I I I Z Z I I\\
        & \qquad\qquad\qquad\qquad\qquad\quad\;-  I I I I I I I I I Z Z I\\
        & \qquad\qquad\qquad\qquad\qquad\quad\;-  I I I I I I I I I I Z Z\\
        & \qquad\qquad\qquad\qquad\qquad\quad\;+  Z I I I I I I I I I I Z\\
    \end{aligned}
    \label{eq:cmaj_paulis_quadratic}
\end{equation}
\setlength{\jot}{3pt}

This can already be a powerful tool for music composition, where the cost function becomes a compositional object, and the behaviour of a specific run influences the result of a given music session (see Sec. \ref{subsec:hexagonal_chambers}).

\paragraph{Chord flips}
\label{subsubsec:chord_flips}

\begin{figure}[ht!]
  \centering
  \begin{subfigure}{.49\textwidth}
    \centering
    \includegraphics[width=\linewidth]{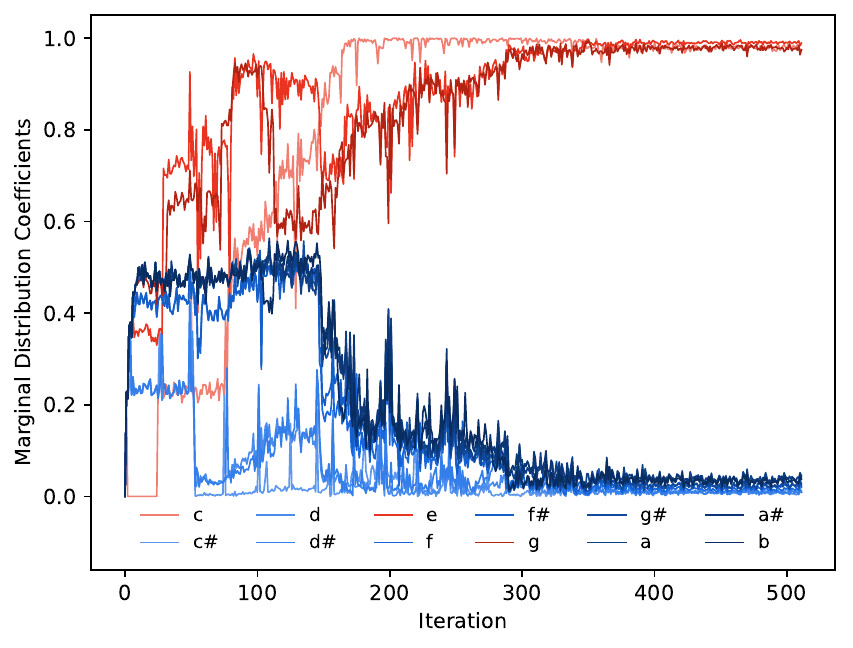}
    \caption{Converging to $C_{maj}$}
    \label{fig:degenerate_a}
  \end{subfigure}
  \begin{subfigure}{.49\textwidth}
    \centering
    \includegraphics[width=\linewidth]{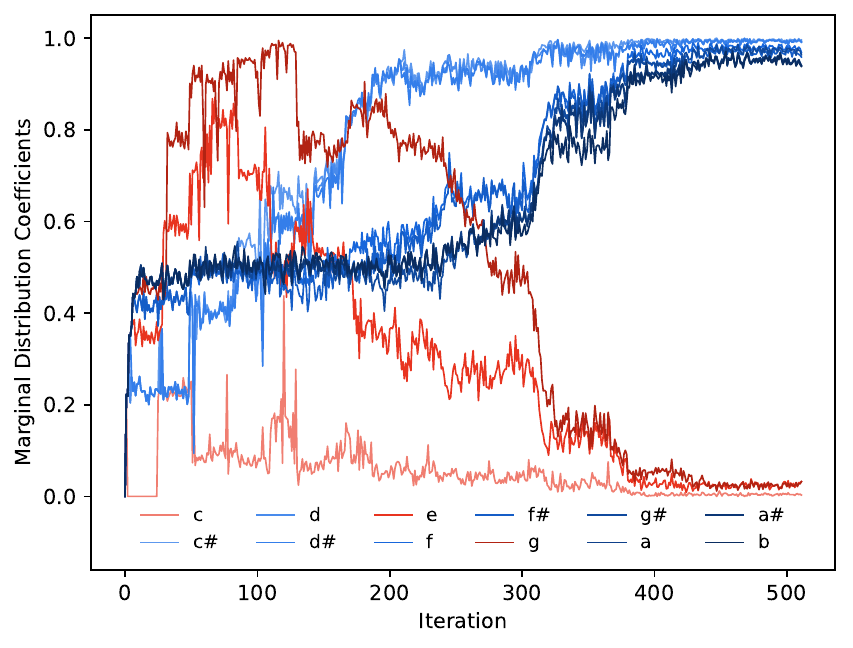}
    \caption{Mixed solution}
    \label{fig:degenerate_b}
  \end{subfigure}
  \begin{subfigure}{.95\textwidth}
    \centering
    \includegraphics[width=\linewidth]{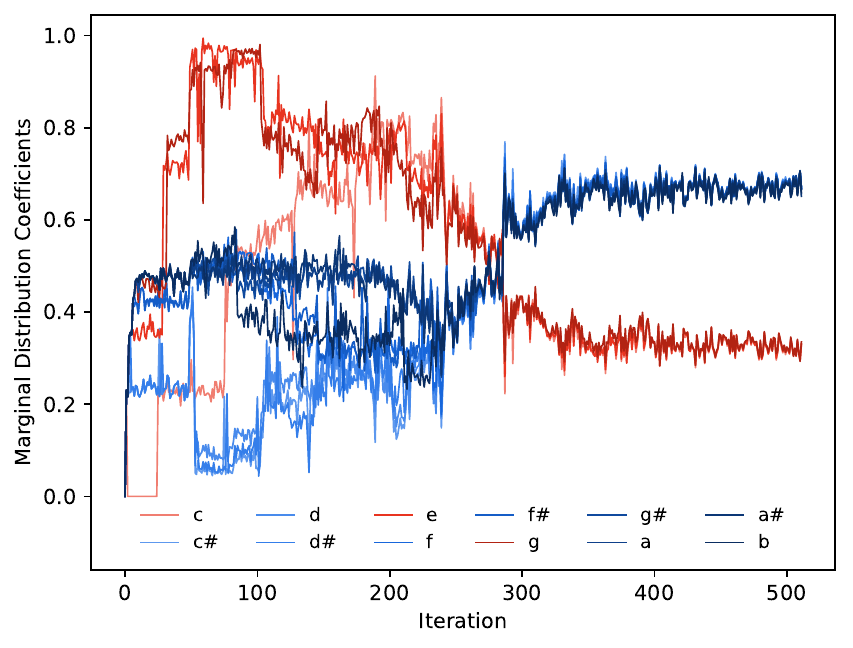}
    \caption{Chord Flip from $C_{maj}$ to $antiC_{maj}$}
    \label{fig:degenerate_c}
  \end{subfigure}
  \caption{Different results for a cost function with degenerate ground state}
  \label{fig:superposition}
\end{figure}

Interestingly, when there is degeneracy in the solution (i.e., more that one state reaches minimum energy), it might happen that occasionally the VQE minimization starts approaching one solution, but suddenly encounters another minima, and switches its trajectory. One example of this behaviour can be seen in \ref{fig:degenerate_c}, where the solution starts by favouring \textit{Cmaj}, but "flips" entirely into an \textit{anti-Cmaj} after around 300 iterations.

\subsubsection{Breaking Degeneracy}
\label{subsubsec:breaking_deg}

Occasionally it might be desirable to consider QUBO coefficients that are highly interconnected, but that still have a single ground state solution. This can be achieved by using corrective linear terms. As an example, starting from equation \ref{eq:ising_cmaj_degenerate}, one could introduce a constant \(\bar{a} \in [-1,1]\) to change the energy symmetry of the system (Eq. \ref{eq:degeneracy_breaking}). In this case, it can be seen that positive values of \(\bar{a}\) will benefit \(C_{maj}\), and vice-versa. \footnote{Notice that this is an empirical result, and not a mathematical proof that \textit{guarantees} symmetry/degeneracy breaking of the ground state.}. 

\begin{equation}
        a_{k} = -\frac{1}{2}\sum_{k\neq l}b_{kl} + \bar{a}
    \label{eq:degeneracy_breaking}\quad\!.
\end{equation}

\begin{figure}[h]
    \centerline{\includegraphics[width=0.7\linewidth]{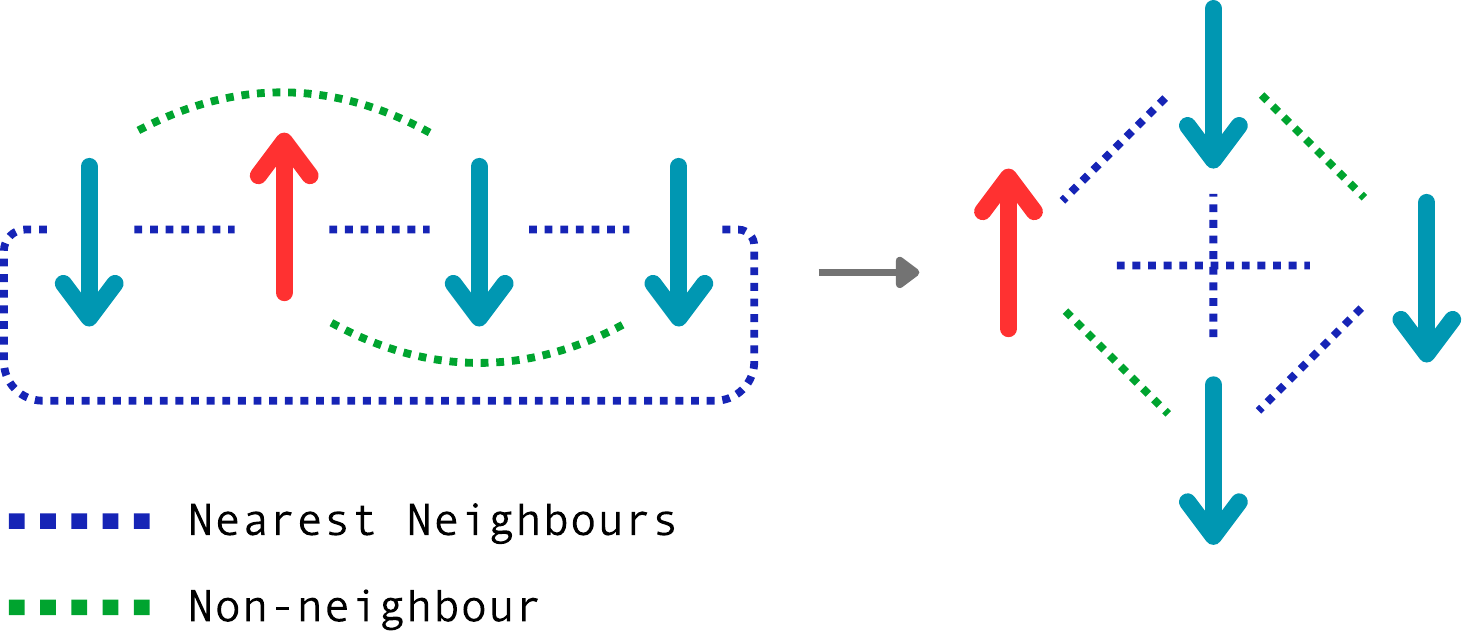}}
    \caption{Increasing the coupling}
    \label{fig:non_neighbour}
\end{figure}

To illustrate, consider the 4-spin example below, where both nearest neighbour and non-neighbour coupling terms are combined to make the second note flip its direction (Fig. \ref{fig:non_neighbour}). Then, a QUBO matrix is designed to satisfy equations \ref{eq:ising_cmaj_degenerate} and \ref{eq:degeneracy_breaking}. Table \ref{tab:4spin} shows the energy spectrum of the system for different values of \(\Bar{a}\). Notice that, as \(\Bar{a}\) increases, the linear solution \(\ket{\downarrow\downarrow\downarrow\downarrow}\) starts to gain importance, until it becomes the predominant term.

\begin{table}[t]
\tbl{Designing QUBOs with linear and quadratic coefficients}
{\begin{tabular}{@{\hspace{2pt}}cc@{\hspace{3pt}}c@{\hspace{3pt}}c@{\hspace{3pt}}c@{\hspace{3pt}}c@{\hspace{3pt}}c@{\hspace{3pt}}c@{\hspace{3pt}}c@{}} \toprule
\multicolumn{9}{c}{\(\langle H\rangle\) (Eqs. \ref{eq:ising_cmaj_degenerate}\ \&\ \ref{eq:degeneracy_breaking})}\\\colrule
\(\Bar{a}\) & \(\ket{\downarrow\downarrow\downarrow\downarrow}\) & \(\ket{\downarrow\downarrow\downarrow\uparrow}\) & \(\ket{\downarrow\downarrow\uparrow\downarrow}\) & \(\ket{\downarrow\downarrow\uparrow\uparrow}\) & \textcolor{blue}{\(\ket{\downarrow\uparrow\downarrow\downarrow}\)} & \(\ket{\downarrow\uparrow\downarrow\uparrow}\) & \(\ket{\downarrow\uparrow\uparrow\downarrow}\) & \(\ket{\downarrow\uparrow\uparrow\uparrow}\)  \\\colrule
4&\textcolor{blue}{-32}&-12&-12&0&-28&0&0&20\\
3&\textcolor{purple}{-24}&-8&-8&0&\textcolor{purple}{-24}&0&0&16\\
1&-8&0&0&0&\textcolor{blue}{-16}&0&0&8\\
0&0&4&4&0&\textcolor{purple}{-12}&0&0&4\\
-1&8&8&8&0&-8&0&0&0\\
-3&24&16&16&0&0&0&0&-8\\
-4&32&20&20&0&4&0&0&-12
\\\colrule
$\Bar{a}$ & $\ket{\uparrow\uparrow\uparrow\uparrow}$ & $\ket{\uparrow\uparrow\uparrow\downarrow}$ & $\ket{\uparrow\uparrow\downarrow\uparrow}$ & $\ket{\uparrow\uparrow\downarrow\downarrow}$ & $\ket{\uparrow\downarrow\uparrow\uparrow}$ & $\ket{\uparrow\downarrow\uparrow\downarrow}$ & $\ket{\uparrow\downarrow\downarrow\uparrow}$ & $\ket{\uparrow\downarrow\downarrow\downarrow}$ \\
\colrule
4 & 32 & 20 & 20 & 0 & 4 & 0 & 0 & -12 \\
3 & 24 & 16 & 16 & 0 & 0 & 0 & 0 & -8 \\
1 & 8 & 8 & 8 & 0 & -8 & 0 & 0 & 0 \\
0 & 0 & 4 & 4 & 0 & \textcolor{purple}{-12} & 0 & 0 & 4 \\
-1 & -8 & 0 & 0 & 0 & \textcolor{blue}{-16} & 0 & 0 & 8 \\
-3 & \textcolor{purple}{-24} & -8 & -8 & 0 & \textcolor{purple}{-24} & 0 & 0 & 16 \\
-4 & \textcolor{blue}{-32} & -12 & -12 & 0 & -28 & 0 & 0 & 20 \\
\botrule
\end{tabular}
}
\begin{tabnote}
$^{*}$Single ground states are denoted in blue.\\
$^{**}$Degenerate Ground states are denoted in purple.
\end{tabnote}
\label{tab:4spin}
\end{table}

\section{VQH implementation}
\label{vqh_implementation}

\paragraph{A Real-Time Approach}A distinctive feature of the implementation of VQH is on its integration, and interfacing between a quantum-centered environment and a music-centered one.

We proposed to realize this communication with a more integrated, modular, and dependency-inverted implementation. The aim is to design the VQH in a way that enables sonifications to be experimented on-the-fly, during a musical performance. Moreover, it could also enable researchers on Quantum Simulation problems to sonify their research while an experiment is running, providing them with a real-time auditory display of an experiment.

To that end, the Variational Quantum Harmonizer became structured as a \textit{modular musical instrument}, or in a broader sense of the word, an \textit{interface for music expression}.

\paragraph{Modularity} The VQH implementation is modular and object-oriented, enabling independent users to define classes in separate python scripts - as specified in the source code documentation \cite{VQHGit} - to add customization and new features, such as different quantum hardware platforms and simulators, complex sonification mappings, new VQA implementations, optimization and simulation problems, encoding/decoding techniques, synthesis engine and music programming language integration.

\subsection{VQH Blueprint}
\label{vqh_blueprint}

\begin{figure}[h!]
    \centerline{\includegraphics[width=\linewidth]{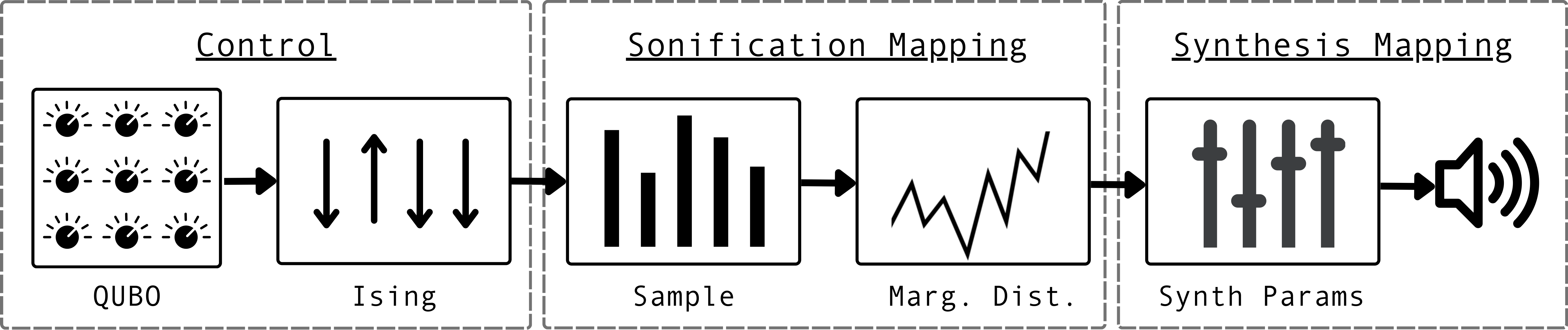}}
    \caption{VQH as a musical interface}
    \label{fig:vqh_stack}
\end{figure}

Based on the classification / characterization system for Digital Musical Instruments proposed by \cite{miranda2006new}, it is possible to represent the VQH by separating the elements related to Control, Mapping and Synthesis\footnote{Note that in a final implementation of a musical interface, there might be no clear separation between, control, mapping and synthesis, as each component might serve in several roles across the system}. In a first layer of this illustration (Fig. \ref{fig:vqh_stack}), it is possible to note that the VQH has one control interface, two mapping stages, and one synthesis engine. In principle, the synthesis side can be implemented in any known synthesis platform or music programming language (See sec. \ref{sec:synth}).

In more detail, the terminology employed in the components of the VQH is explained: When considering a general VQA algorithm, the main input of the system is the Hamiltonian of a system (\textit{\textbf{Problem}}), where the energy expectation values to be minimized are extracted from (\textbf{\textit{Control}}). This Hamiltonian is described as a list of operators decomposed as Pauli tensor products. The first mapping stage deals with polling data from the iterations of the VQA (\textbf{\textit{Generation}}), which could be running with a QASM Simulator or in a real device (\textbf{\textit{Platform}}) and decoding the information that will be used as a data stream for sonification (\textit{\textbf{Protocol}}). Then, this data is transformed in the next stage into specific synthesis parameters, according to the chosen sonification strategy (\textit{\textbf{Mapping}}). Finally, a synthesizer listens to these parameter streams and plays sounds (\textit{\textbf{Synthesis}}).

Additionally, there is a higher level interface, where the user controls more global parameters of the system, general VQE configurations, editing parameters and variables, as well as general controls for running experiments and triggering sound (\textbf{\textit{Interface}}). Further, there is a lower-level control, where it is possible to design the problem that is being optimized (e.g., the Pauli operators) and switch between the sonification mapping strategies, as well as choose how the sonification data is read and parsed to a synthesis engine (\textbf{\textit{Core}}).

\begin{figure}[ht!]
    \centerline{\includegraphics[width=0.85\linewidth]{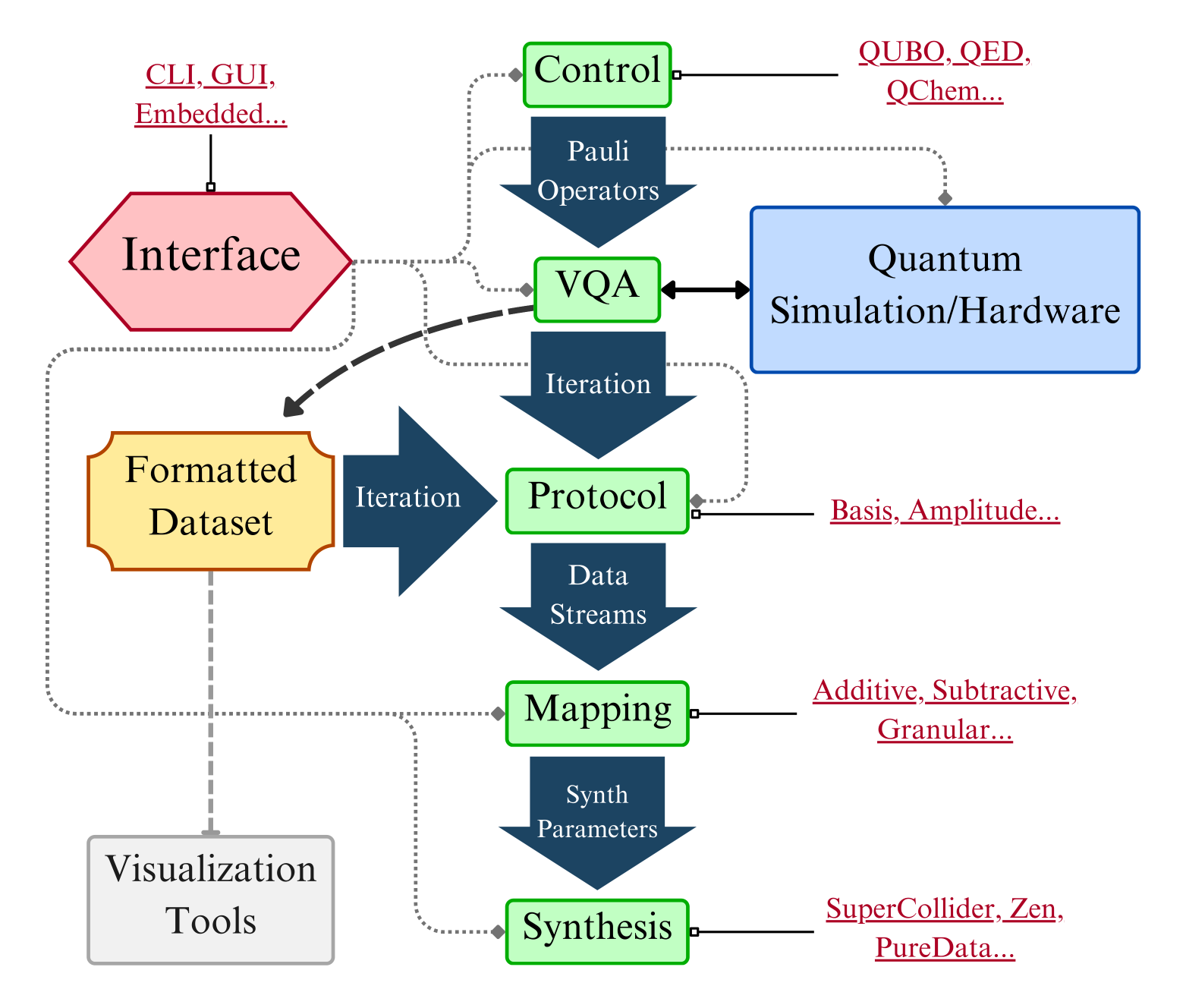}}
    \caption{Generalized VQH Structure}
    \label{fig:vqh_components}
\end{figure}

A visualization of the implementation can be seen in the diagram below (Fig. \ref{fig:vqh_components}). Note that a box was inserted in the diagram to indicate that the data can also be plotted and visualized as an image by an external software (see Sec. \ref{subsubsec:hc_sync_vqh} and Fig. \ref{fig:vqh_zen} for an example). Generally, it is a conjunction of audio and image (and other sensorial media) that constitute the global concept of data visualisation, compehension and new insight on a dataset. In so, the VQH aims to be flexible for the introduction of other visualization techniques apart from sound.

\subsection{Control}
\label{subsec:control}

\subsubsection{Controlling the QUBO Problem}
\label{subsubsec:control_qubo}

As seen in Sec. \ref{sec:vqh_overview}, the main problem studied so far within the VQH is the QUBO problem. The QUBO is transformed into an Ising Hamiltonian, by changing the binary variable basis \(n_i\), which has values \(0\) and \(1\), into a Pauli \(Z\) basis, having eigenvalues \(\pm 1\). In the previous section \ref{sec:designing_chords}, it was shown how the user can design cost functions using the harp analogy and interact with linear and quadratic QUBO coefficients to obtain specific results. When more complex couplings are used, knowing the ground state of your problem becomes more difficult.

\paragraph{Creating Superposition} More generally, we can consider adding a transverse magnetic field to our Hamiltonian, controlled by a magnitude parameter \(h_x\), which can introduce phase transitions and quantum effects to the Ising model (Eq. \ref{eq:transversefieldising})\footnote{This parameter was also used for creating a musical gesture for \textit{hexagonal Chambers} (Sec. \ref{subsec:hexagonal_chambers}), where the transverse field increased progressively towards the end of the piece}.

\begin{equation}
H(\sigma)=\sum_i \alpha_i \sigma^Z_i + \sum_{i,j\neq i} \beta_{ij}\sigma^Z_i \sigma^Z_j - h_x \sum_i \sigma^X_i
\label{eq:transversefieldising}
\end{equation}

The introduction of a transverse field can lead to systems in which the ground state is a superposition state. As a result, with a sufficiently large field, it can be seen from equation \ref{eq:transversefieldising} that the ground state of the Ising problem may be alternatively written in terms of the \(\sigma^X\) basis (\(\ket{\rightarrow}\),\(\ket{\leftarrow}\)). In this case, a different mapping protocol could be designed, where measurements in the \(\sigma^X\) basis are also considered.

\begin{figure}[h]
    \centerline{\includegraphics[width=0.67\linewidth]{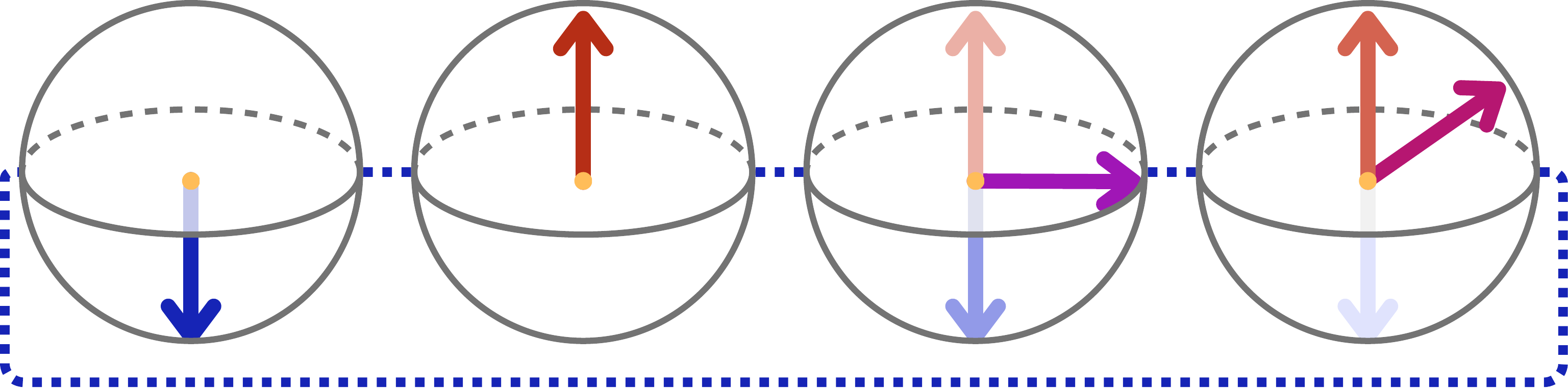}}
    \caption{Creating Superposition with transverse fields}
    \label{fig:transverse_ising}
\end{figure}

\subsubsection{Controlling the VQE}
\label{subsubsec:control_vqe}

One of the most relevant components of a VQA is the classical optimizer. As seen in \ref{par:vqe}, while the quantum side of the algorithm is responsible for evaluating the energies for a given parameterized state, the classical optimizer is the element that guides the changes in the configuration space. Thus, changing the optimizer can have a significant impact in the trajectory towards finding a result. Since the VQH is sensitive to iteration-by-iteration data, this also translates into noticeable perceptual changes in the sonification. 

\begin{figure}[h!]
\centering
  \begin{subfigure}{.48\textwidth}
    \centering
    \includegraphics[width=\linewidth]{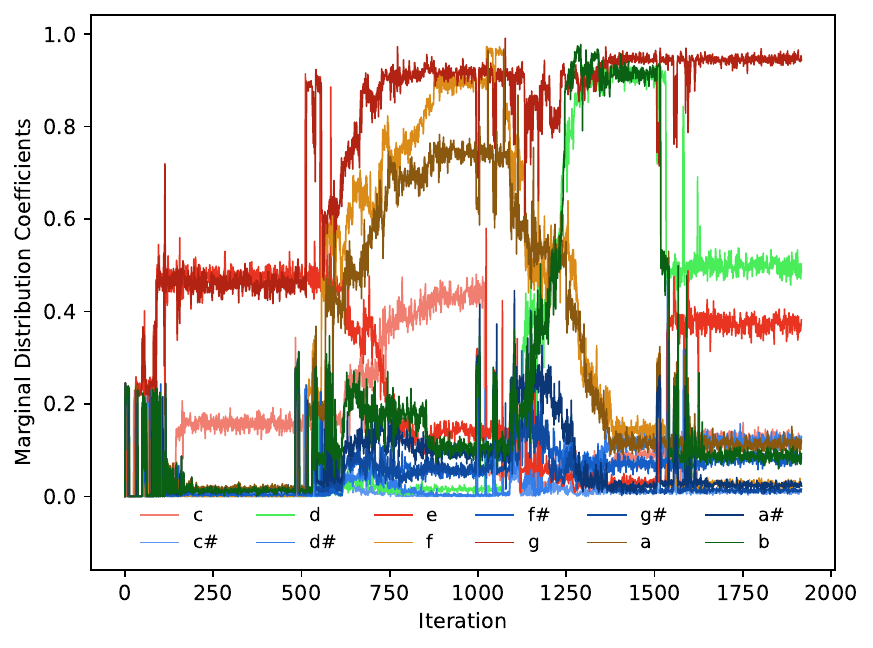}
    \caption{COBYLA}
    \label{fig:cobyla}
  \end{subfigure}
  \begin{subfigure}{.48\textwidth}
    \centering
    \includegraphics[width=\linewidth]{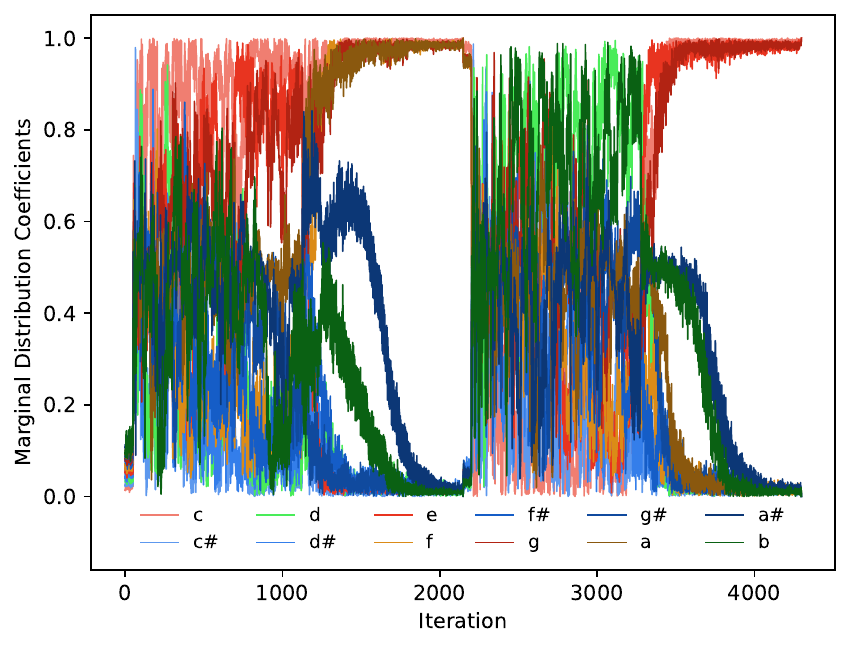}
    \caption{SPSA}
    \label{fig:spsa}
  \end{subfigure}
  \begin{subfigure}{.48\textwidth}
    \centering
    \includegraphics[width=\linewidth]{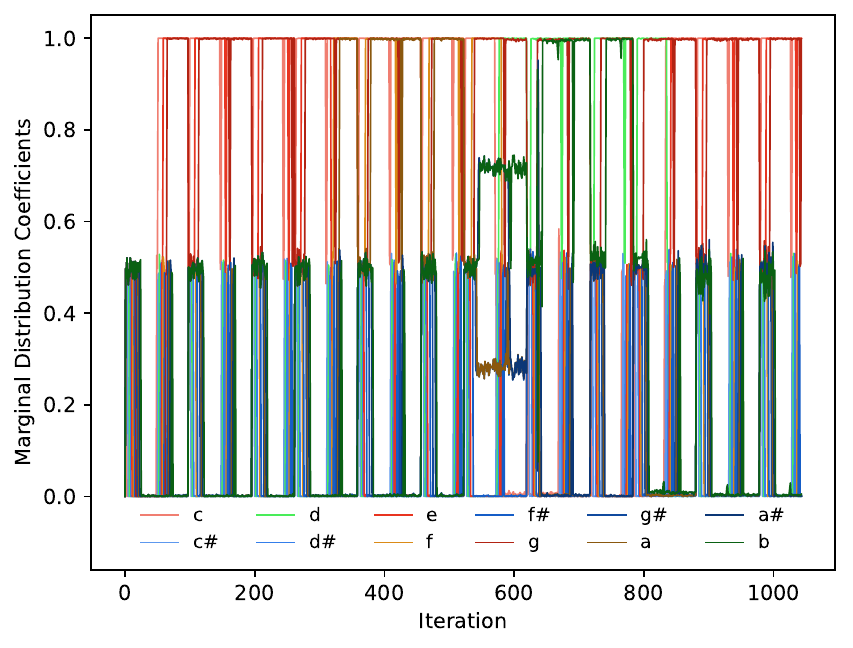}
    \caption{NFT}
    \label{fig:nft}
  \end{subfigure}
    \caption{Comparison between Optimizers}
    \label{fig:optimizers}
\end{figure}

\paragraph{Artistic Intuition on the Classical Optimizers}
In Fig. \ref{fig:optimizers}, a comparison between three optimizers is made for a chord progression (Sec. \ref{subsec:chord_progressions}). Restating the interpretation mentioned in \cite{ISQCMC_itaborai}, the COBYLA algorithm (\ref{fig:cobyla}) seems to ``perturb" each note individually while analyzing the changes in the cost function. Using the harp analogy for the basis encoding, it looks like the algorithm is ``sweeping" through the strings (or muffling them), while deciding which tones ``sound better" according to our hamiltonian listener. 

On the other hand, SPSA \cite{spsa1992} (Fig. \ref{fig:spsa}), has a more noisy profile, where the marginal distributions show a slower convergence. This means that usually many notes could be sounding simultaneously, until non-desired notes progressively fade out. This case could be particularly useful for constrained-randomness applications in composition, where parameters vary within moving boundaries during a complex gesture.

The NFT \cite{nakanishi2020sequential} algorithm (\ref{fig:nft}), in contrast, generates a periodic pattern, which is useful for creating rhythmic pulses.

\subsubsection{Implementation}
\label{subsubsec:vqh_files}

\begin{figure}[ht!]
    \centering
    \includegraphics[width=0.67\linewidth]{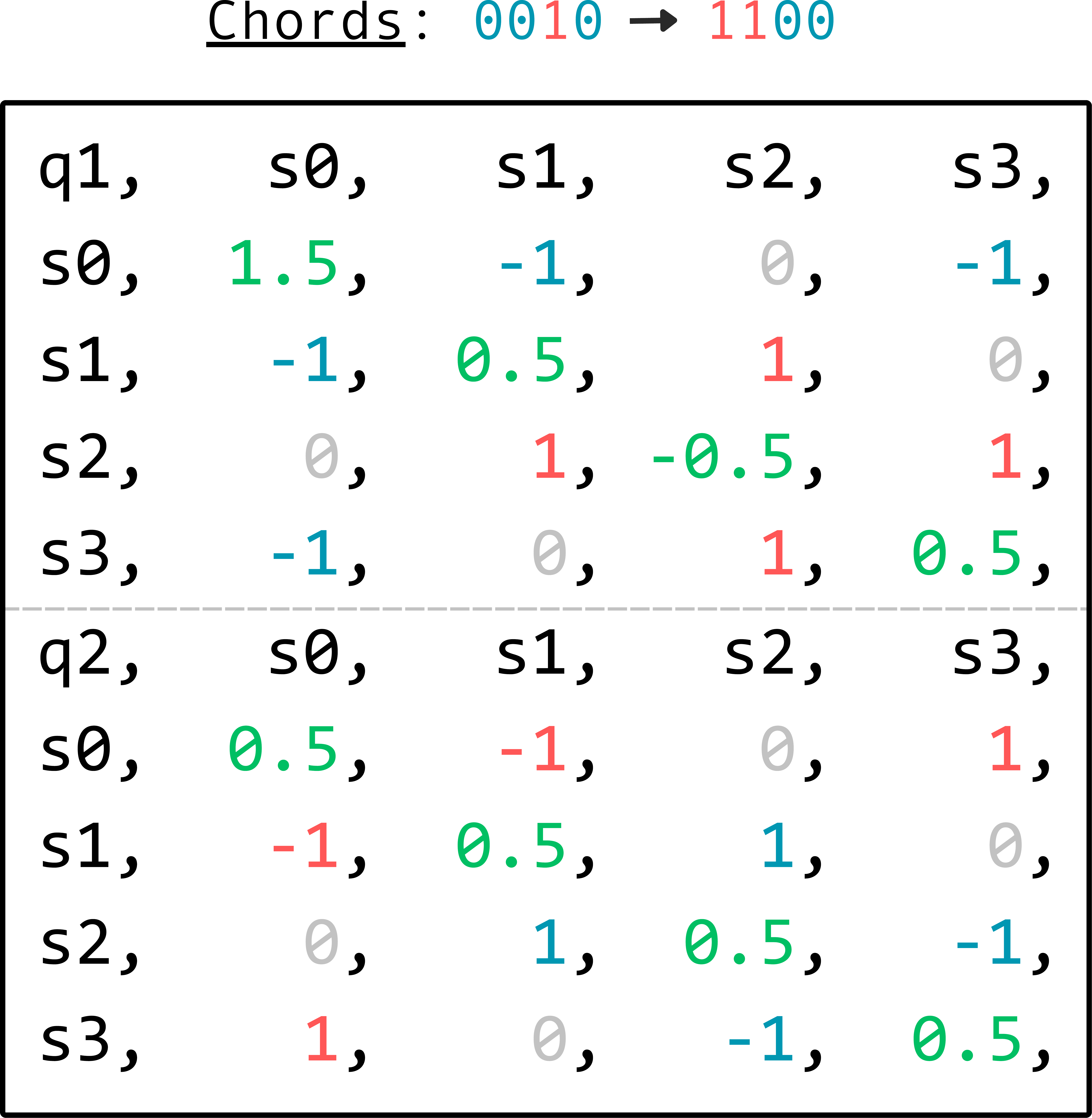}
    \caption{Text-Based control of a chord transition (Eq. \ref{eq:degeneracy_breaking})}
    \label{fig:h_setup}
\end{figure}

\paragraph{QUBO Matrix} In the VQH software, the QUBO coefficients are stored and read from a comma separated matrix form in a CSV file. As shown in Figure \ref{fig:h_setup}, there is a header line that provides labels and metadata to build the pauli operators and to facilitate the sonification mapping.
To facilitate the creation of chord progressions, it is possible to define several Hamiltonians in the CSV file, in sequence. Figure \ref{fig:h_setup} shows a 4-qubit example of a chord transition, using Eq. \ref{eq:degeneracy_breaking} with \(\Bar{a} = 0.5\).

\paragraph{VQE Configuration} Other relevant configurations for the VQE experiment are found in a separate JSON file (Fig. \ref{tab:vqe_conf}). If there is more than one QUBO matrix stored in the file (\texttt{sequence\_length}), the VQH will parse the result of one experiment as the initial point for the next.

\begin{table}[h]
\tbl{VQE configuration file on current version of VQH}
{\begin{tabular}{@{}cc@{}} \toprule
Argument Name & Argument Description \\\colrule
\texttt{reps} & Number of Layers for the Ansatz \\
\texttt{entanglement} & Type of Entaglement$^{*}$ \\
\texttt{optimizer\_name} & Classical Optimizer \\
\texttt{sequence\_length} & Number of Hamiltonians \\
\texttt{size} & Size of the problem in qubits\\
\texttt{description} & Experiment Metadata\\
\texttt{iterations} & VQE iterations per Hamiltonian (list)\\
\texttt{nextpathid} & Experiment's unique idenfifier\\\botrule
\end{tabular}
}
\begin{tabnote}
$^{*}$Specific for the EfiicientSU2 Ansatz.
\end{tabnote}
\label{tab:vqe_conf}
\end{table}

\subsubsection{Text-based and MIDI-based control}
\label{subsubsec:control_text_midi}

Initially, the user could control the QUBO coefficients by directly editing and saving the CSV file using a preferred text editor. Similarly, the VQE configuration can be edited manually.

This approach can be useful for \textit{Live Coding} performances (see Sec. \ref{subsec:live_coding}), where typically there is an aesthetic demand for displaying text editors to the audience, sharing how the programmer-artist implements the sonification and interacts artistically with the code during the performance.

\begin{figure}[h]
    \centerline{\includegraphics[width=0.6\linewidth]{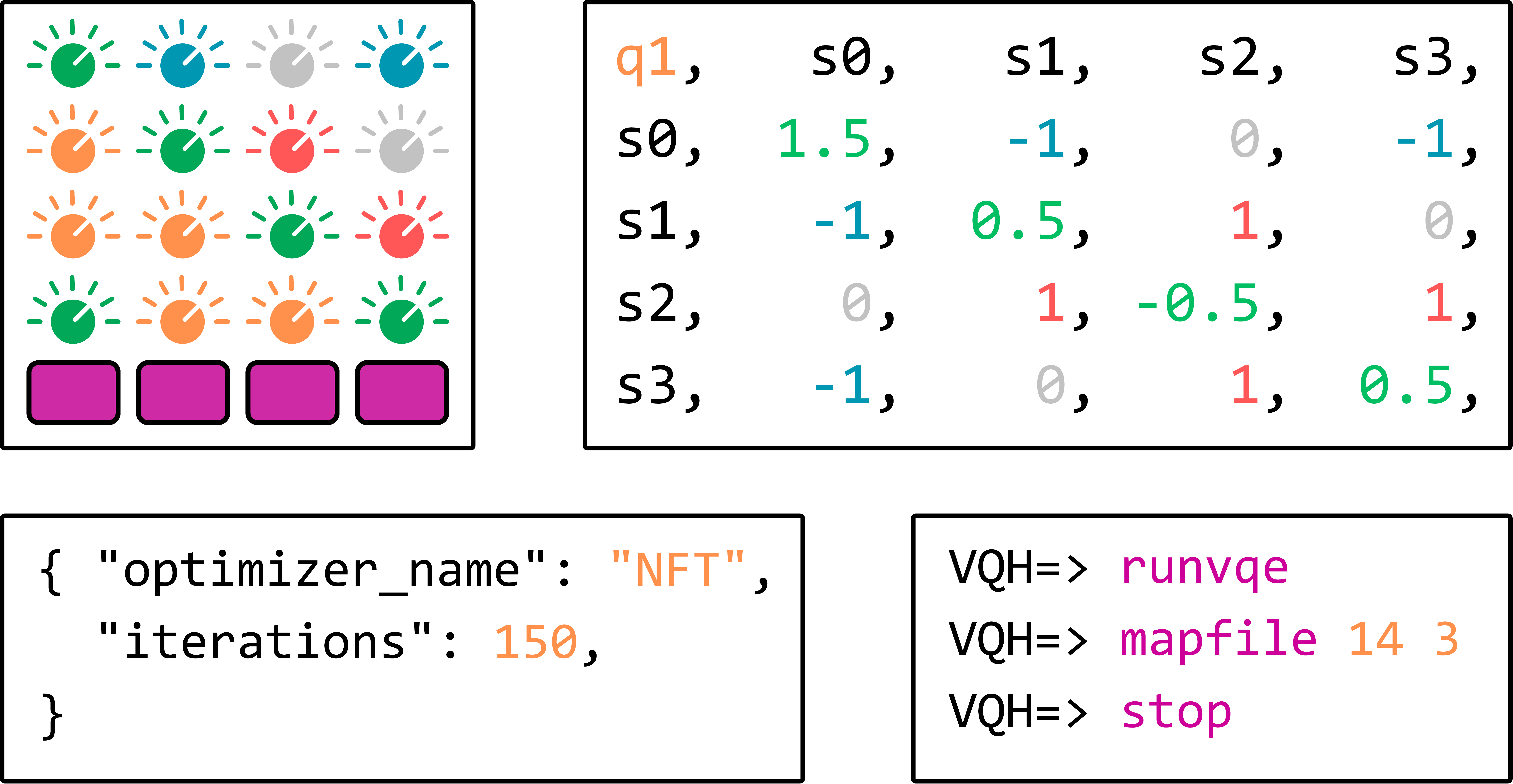}}
    \caption{Example of a MIDI Controller assigned to VQH parameters}
    \label{fig:control_midi_txt}
\end{figure}

Extending this control, it is also possible to assign MIDI Control Change values in VQH to indirectly control the QUBO coefficients and VQE parameters.
For instance, one could use a MIDI knob grid to change coefficients and a button/pad matrix to control configuration settings, change the mapping and toggle the sounds (Fig. \ref{fig:control_midi_txt}).

\subsection{CLI Interface}
\label{subsec:cli}

The current VQH version can be operated through a Command-Line Interface, implemented as an encapsulated prompt environment, with pre-defined functions for managing the interfacing between all the components of the instrument.

Inside the VQH file structure and a python environment with all dependencies installed, it is possible to run an \textit{Experiment Session} by running the main script \texttt{VQH.py} and selecting the script arguments, as described below.

Within a specific \textit{Session}, the user can encapsulate a series of experiments, metadata and output data in a dedicated folder, which can be retrieved and used at a later time.

\subsubsection{Script Arguments}
\label{subsubsec:cli_args}

\begin{verbatim}
python VQH.py [SESSIONPATH] [PLATFORM] [PROTOCOL]
\end{verbatim}

\begin{itemize}
	\item \texttt{SESSIONPATH} - The name of the folder where the VQE experiments and all generated data will be stored for this session. If the folder already exists, more data will be appended to the same folder.
	\item \texttt{PLATFORM} - Quantum Provider used. This is where the connection to quantum computing services is made possible. If a user has access to cloud services, their credentials can be used to connect to remote providers and execute jobs there. Alternatively, a local simulator can be used. 
	\item \texttt{PROTOCOL} - Note encoding schemes/protocols for interpreting quantum states as notes, determining the global layer of sonification mapping. At the time of writing, the only implemented protocol was the \texttt{basis} protocol (See Sec. \ref{subsubsec:harp_analogy}).
\end{itemize}

\paragraph{Example}

\begin{verbatim}
$ python VQH.py Example local basis
\end{verbatim}

In the session defined above, all generated data will be stored inside a folder named \texttt{Example\_Data/}, each experiment has its own unique identifier '\texttt{<id>}', and its respective results will be found in the subfolder \texttt{Example\_Data/Data\_<id>}. Moreover, the session will perform jobs using a local provider (\texttt{qiskit\_aer}).

\subsubsection{VQH Prompt}
\label{subsubsec:vqh_prompt}

After startup, the user is presented with a prompt "\texttt{VQH=>}", where internal commands can be used for operating the sonification.

\begin{itemize}
	\item \texttt{VQH=> runvqe} - This command triggers a VQE experiment. When called, the command reads the current configuration files, the protocol being used, as well as the QUBO coefficients defined in the \texttt{h\_setup.csv} file. Then the minimization algorithm is executed, following the specification, and the sonification data is saved in a folder with an assigned identifier '\texttt{<id>}'. If more than one matrix was defined, it will run VQE for each Hamiltonian using the previous results as initial point, and concatenating them for saving the final dataset.
	\item \texttt{map [type]} - This command triggers the sonification for the last experiment that was run and loaded into memory. The [type] flag denotes the sonification method and synthesis engine used (see Sections \ref{sec:mapping} and \ref{sec:synth}).
	\item \texttt{mapfile [<id>] [type]} - Triggers the sonification using data from a previous experiment stored in the session folder, using the \texttt{<id>} to select the dataset.
	\item \texttt{stop} - Sends a ``panic" message to stop all sounds. Necessary for some sonification mappings where the generated sound is continuous.
	\item \texttt{quit} or just \texttt{q} - Exits the program.
\end{itemize}

\section{Sonification Mapping}
\label{sec:mapping}

After exploring the VQH implementation, and exploring different cost function designs and how the datasets are generated using the VQH interface, the focus is switched for the sound generation. This is a typical sonification mapping problem.

\subsection{Sonification Data}
\label{subsec:son_data}

At this stage of the VQH pipeline, it will be considered that an experiment session was run, and a dataset was generated and stored in a folder. Inside this folder, a set of files can be found.

Firstly, there are three files that contain metadata for the reproduction of the experiment. Respectively, they contain a copy of the QUBO coefficients, the VQE parameters, and a list of operators used. Secondly, there is a \texttt{.json} file containing raw results - a list of sampled distributions for each iteration. Third, there are files containing post-processed according to the chosen protocol, such as marginal distributions, observables, amplitudes, etc\footnote{Notice that since the raw results are saved, the user can retrieve a previous experiment and sonify it using a different protocol.}.

More specifically, after using the basis protocol, there should be three files of post-processed data:

\begin{itemize}
    \item A stream of Marginal Distribution Coefficients of a sampled statevector for each iteration (\(c_n(t)\)).
    \item Energy Expectation Value for each iteration (\(E_0(t)\)).
    \item The basis state with the highest sampled probability. (\(\ket{\psi_{max}(t)}\)) 
\end{itemize}

Then, the mappings described below were proposed to map the data streams into synthesis control parameters.

\subsection{Simple Mappings}
\label{subsec:simple_maps}

\subsubsection{Additive}
\label{subsubsec:additive}

One of the most direct and natural approaches one thinks about when considering a mapping problem is to use the simplest constituents of sound: Amplitude and frequency. 

Subsequently, it is possible to consider the MDC as an ideal candidate for controlling relative amplitudes (or defining frequency variation). Furthermore, this framework is directly correlated with the harp analogy presented previously \ref{subsubsec:harp_analogy}, where each qubit is assigned to an individual note. In addition, there are a myriad of options when it comes to the choice of frequencies for each note, which can fall into any desired musical scale.

More generally, any method that consists of adding fundamental sounds, such as pure tones, oscillators, and sound banks, to create notes, chords, timbres, clusters and noises is considered to be \textit{additive synthesis}. As a result, controlling the relative amplitudes of these sounds can change the desired timbre and characteristics of the perceived sound.

\subsubsection{Frequency Modulation}
\label{subsubsec:fm}

Under a similar additive structure, it is possible to use the MDC streams to control frequency variation. For each coefficient \(c_n(t)\), there is a frequency \(f_n(t)\).

In a preliminary approach, one could consider a linear (or logarithmic) mapping between the minimum and maximum values for the coefficients (\(c_{min}, c_{max}\)) and an audible range of frequencies (\(f_{min}, f_{max}\)), as exemplified below.

\begin{align}
    &f_n(t) = \frac{(f_{max}-f_{min})}{(c_{max} - c_{min})}(c_n(t) - c_{min}) + f_{min}\;\; (\text{linear})\\
    &f_n(t) = f_{min}.\left(\frac{f_{max}}{f_{min}}\right)^{\frac{(c_n(t)-c_{min})}{(c_{max} - c_{min})}}\;\; (\text{logarithmic})
\end{align}

\paragraph{Inharmonicity Approach} Another FM approach to be considered for sonification is proposed. Consider a tone generated from a harmonic series. For a given partial \(n\), with frequency \(nf_0\) it is possible to define a proportional shift $c_n(t)$ away from the harmonic equilibrium. In other words, it introduces inharmonicities to the sounds, which can potentially create complex metallic timbres (Eq. \ref{eq:inharmonic}). 

\begin{equation}
    f_n(t) = (n - \frac{(c_n(t)-c_{min})}{(c_{max} - c_{min})})f_{0}
    \label{eq:inharmonic}
\end{equation}

\subsubsection{Subtractive}
\label{subsubsec:subtractive}

Instead of accumulating tones, an alternative method could be to employ a subtractive or filtering strategy. One could generate a white noise (or start from any spectrally rich sound), and filter it down using a parallel set of Resonant (Band-Pass) Filters (RF).

Similar to the additive approach, it is possible to center each RF in frequencies that follow a determined musical scale. Then, each component of the MDC (\(c_n(t)\)) can be assigned as the gains of each filter.

To increase the complexity of the generated sound, the energy expectation value (\(E_0(t)\)) will be included, to control a second parameter of the synthesis. For instance, it could control the RF quality/resonant factor \(Q\). A low value of energy could represent a high \(Q\), and vice-versa. As a result, as the expectation value decreases, the sound approximates a pure tone, inferring a perceptual result similar to the harp analogy. Conversely, as the result diverges from the ground state, it approaches the original sound.

\begin{figure}[ht!]
    \centering
    \includegraphics[width=0.8\textwidth]{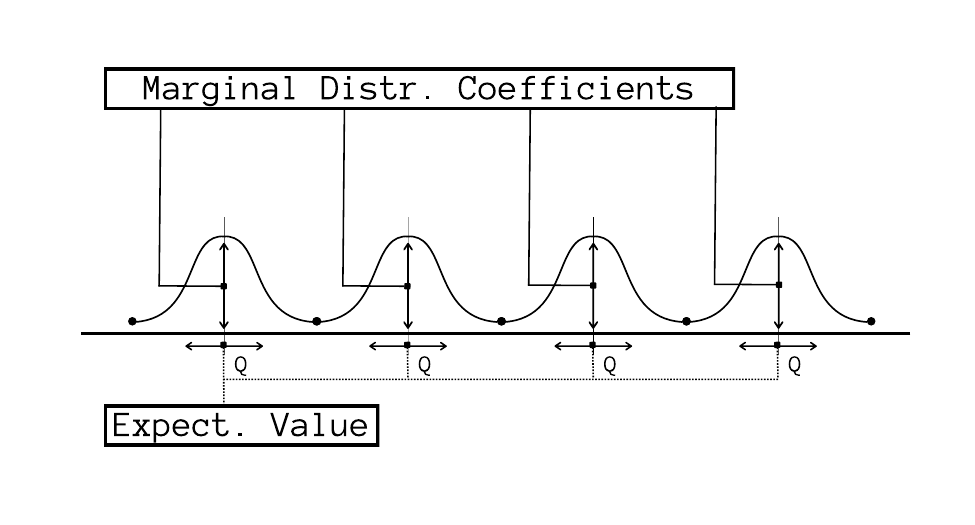}
    \caption{Subtractive Synthesis Mapping}
    \label{fig:subtractive}
\end{figure}

\paragraph{Rotary motors}
An alternate mapping for the expectation values, being a one-dimensional control signal, would be to leave apply a global pitchshift to the sounds - in other words, modulate the RF central frequencies. Depending on the intensity of the modulation and the RF resonant factor, the perceptual result can resemble a rotary motor experiencing variable friction, such as a drilling machine.

\subsection{Intermediary Mappings}
\label{subsec:intermediary_maps}

\subsubsection{Arpeggios}
\label{subsubsec:arpeggios}

For some applications, it might be desirable to listen to each marginal distribution coefficient more individually for a different distinction between relative amplitudes. This can be achieved by designing a \textit{temporal expansion} around a VQE iteration.
For example, instead of listening to a chord or cluster, the sound could be distributed in time as an \textit{arpeggio}. 

A suggested method for arpeggiation is demonstrated. First, in a fixed iteration \(i\) the collected coefficients \(c_n(i)\) are mapped as intensities of individual percussive notes \(c_n(i)p_n\). Then, the notes are sorted by amplitude, from the quietest to loudest note. These notes are played in a rapid succession, achieving an arpeggio. 
In addition, the time expansion rate (speed) of the arpeggio could be controlled by the expectation value. The closer from the ground state, the more compact the arpeggio becomes, until all notes ultimately become simultaneous again.

\begin{figure}[ht!]
    \centering
    \includegraphics[width=.8\textwidth]{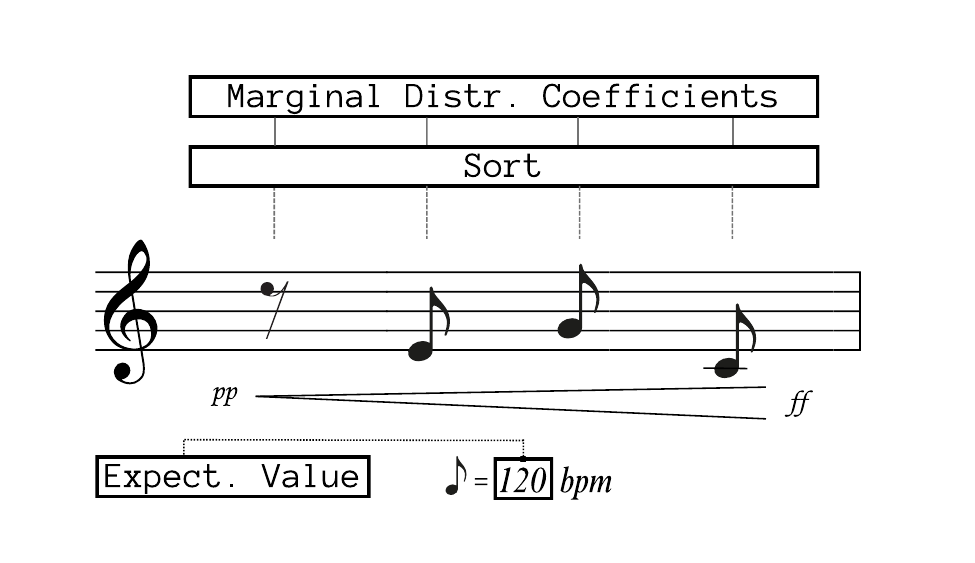}
    \caption{Arpeggio Mapping}
    \label{fig:arpeggio}
\end{figure}

\subsubsection{Spatialization / Diffusion}
\label{subsubsec:spat_diffusion}

Another perceptual attribute of sound often used in music composition is its spatial location.
Stereophony, for example, is popularly used in the recording industry. However, there are also many compositions that carefully position sound objects in space, using surround systems. Among the algorithms used to distribute sounds in space, we highlight Vector-Based Amplitude Panning (VBAP)\cite{pulkki1997virtual}, Distance-Based Amplitude Panning (DBAP)\cite{lossius2009dbap}, and higher-order Ambisonics \cite{poletti2000unified}. 

\begin{figure}[h]
    \centerline{\includegraphics[width=0.7\linewidth]{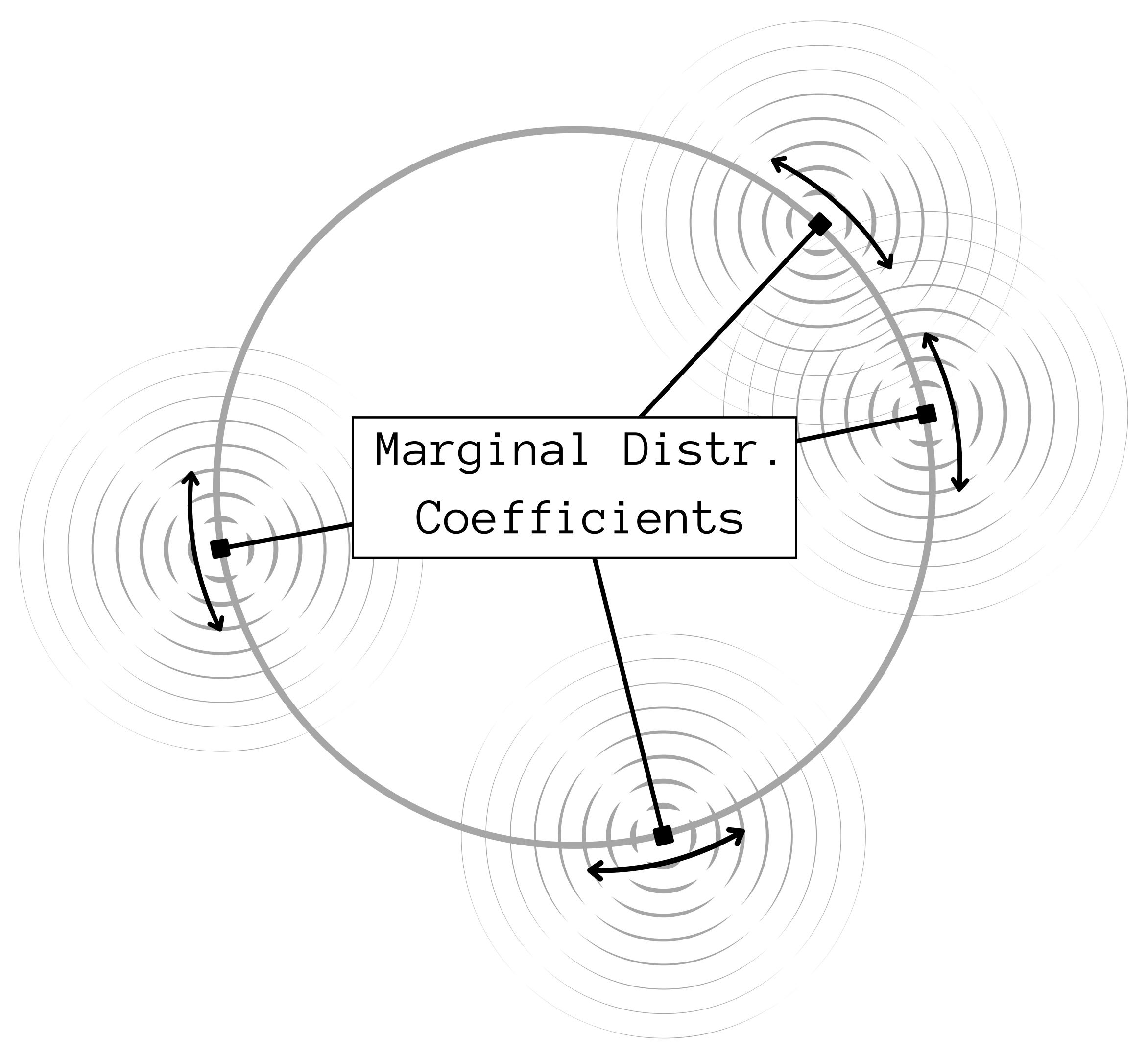}}
    \caption{Spatialization Mapping}
    \label{fig:spat}
\end{figure}

In contemporary music, sound systems are typically arranged in a circle around the audience, enabling the diffusion of quadraphonic, hexaphonic, octaphonic or hexadecaphonic sounds. The latter is usually configured in an upper and lower ring with 8 speakers each, achieving a three-dimensional spatialization. 
For illustration purposes, consider a sound object that can be placed at a fixed virtual distance away from a listener in any direction (Fig. \ref{fig:spat}). In this way, its position can be characterized by a parameter \(\phi(t)\), which describes a dynamics around the circle. If there are $N$ sound objects, their positions could be controlled by the MDC.

\begin{equation}
    \phi_n(t) = 2\pi\frac{(c_n(t)-c_{min})}{(c_{max} - c_{min})}
\end{equation}

\paragraph{Spectral Diffusion}
An interesting example of spatialization consists of dividing a sound into frequency bands using filter banks and spreading each band around the space \(\phi_n\), resulting in a spectral diffusion.

\subsubsection{Granular Synthesis}
\label{subsubsec:granular}

Granular synthesis explores the phenomenon of sound in terms of particles rather than waves. Sparse sonic occurrences, or grains, are perceived by the ear as discrete, rhythmic events. As the grain density increases, a perceptual continuity emerges that can be experienced as pitch \cite{roads2004microsound}. Grains can be short segments from a variety of sources - deterministic waveforms, noise or samples - and can be shaped with a variety of envelopes.

The sonic parameters of each grain can be considered within two perceptual domains simultaneously: temporally, including grain duration and amplitude envelope, and spectrally, encompassing the frequency content of the individual grain and its perceived pitch. Moreover, the holistic organisation of grains can be approached in a number of ways, such as the granulation of prerecorded sounds or the implementation of physical models\cite{roads2004microsound}.

The wide range of time-domain and frequency-domain synthesis parameters used to define grains, along with the numerous algorithms employed to organise them, offers a wealth of opportunities for exploring different sonification methodologies within granular synthesis. We shall explore some of these techniques while presenting the compositional outcomes of the VQH in a following section (\ref{sec:composing}).

\begin{figure}[h]
    \centerline{\includegraphics[width=0.8\linewidth]{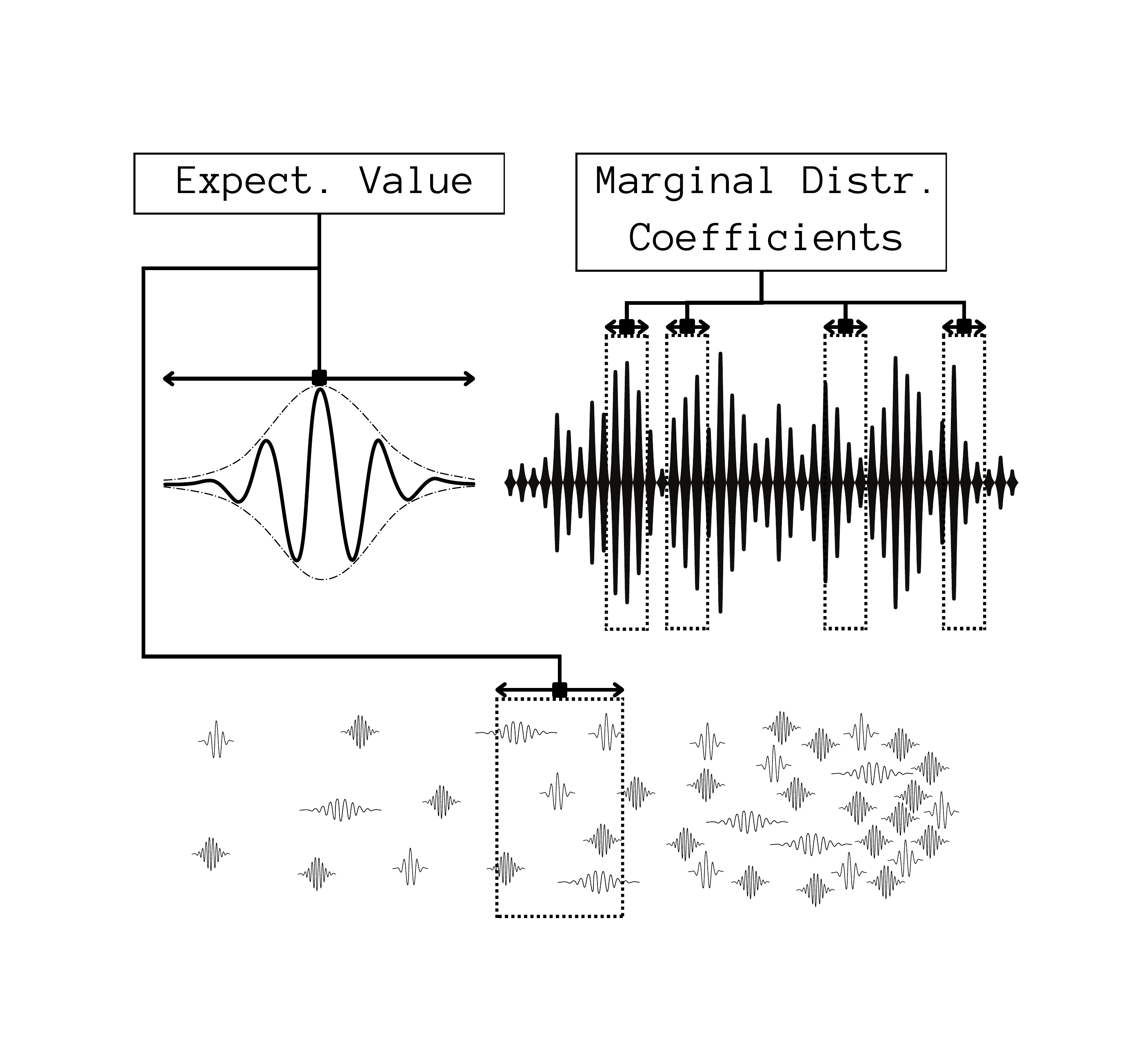}}
    \caption{Granular Synthesis Mapping Example}
    \label{fig:granular}
\end{figure}

\subsection{Advanced Mappings: Live Coding}
\label{subsec:live_coding}

Several of the mapping methods discussed previously operate within a linear input-output framework. Here, the results obtained from the Variational Quantum Eigensolver serve as control data for the chosen sonification strategy, which then yields a single output. This approach aligns with more traditional methods of algorithmic composition; computational constraints require the composer to prepare a script in advance and then wait for the computer to synthesize the results.

With the advancement of computational power and the implementation of just-in-time programming within musical languages (as seen in Supercollider's JITLib), a seamless real-time interaction between composer and machine became achievable. The emergence of concepts such as ``conversational programming," ``on-the-fly programming," or ``interactive programming" has led to the development of numerous creative languages and practices. Commonly termed 'live coding,' these activities have reshaped the landscape of algorithmic music over the past decades, moving it beyond its traditional academic confines to reach a wider audience \cite{magnussonmclean2018}. Present-day live coding performances are hosted in a variety of both academic and non-academic contexts, where programmers craft code on stage while their laptop screens are projected for the audience to observe.

By utilising an appropriate live coding system, our aim is to engage with the data yielded by the VQE in an exploratory manner. This will make possible the concurrent employment of multiple synthesis techniques, as well as the ability to dynamically adjust the compositional algorithm in real-time as the data stimulates the creative ideas of the performer. In other words, in Live Coding, we depart from a more direct mapping of data onto sonic parameters, and move towards the use of data as a catalyst for creativity.

\section{Synthesis}
\label{sec:synth}

The implementation of a mapping strategy depends greatly on the synthesis platform used and how it can be connected to Python. The authors chose two platforms to carry the current work. First, Supercollider, which was used to implement the simple and intermediary mapping strategies. Then, the advanced mappings were implemented using Zen.

\subsection{Supercollider}
\label{subsec:supercollider}

To establish a communication between the two systems, a Python client for SuperCollider was used \cite{pythonsupercollider}. This package implements API messages to carry messages to a supercollider server though OSC messaging, eabling us to instantiate synthesizers, flexibly change synthesis parameters, manage nodes and audio buses, load buffers and trigger patterns from python. Altough it does not implement a fully featured \texttt{sclang}, it suffices for the purposes of this work.

The methodology is as follows. First, the user creates and compiles a synthesizer in SuperCollider, using the \texttt{SynthDef} Unit Generator. All parameters that should be controlled by python should be exposed as induvidual function arguments. Then, by using the "\texttt{.store}", method, the synthdef is compiled into machine code and saved automatically in a system folder.

Thereby, the key used to name the syntheziser can be used to create a synth node from python. Further, all exposed attributes can also be controlled by a '\texttt{.set()}' message.

\subsection{Zen}
\label{subsec:zen}

Zen is a live coding language designed for expressing multidimensional musical patterns through succinct code, with a particular focus on pattern interference. Beyond the language itself, the Zen ecosystem encompasses an integrated development environment, featuring a code editor, pattern visualiser, and synthesis engine. All components are developed for the web, creating a comprehensive performance tool that necessitates no installation beyond a modern browser. Zen is built upon JavaScript, offering a familiar foundation for the majority of artist-programmers. However, it also incorporates some domain-specific syntax to augment JavaScript's capabilities.

Through the textual interface, Zen allows the user to express causal relationships between musical layers. A composer is able to define discrete patterns, and then identify musical or sonic parameters between patterns that should interfere with each other; for example, pitch, amplitude, timbral, or spatial parameters. Any parameter from another musical layer can be utilised and manipulated as the input for any other parameter. Composing in this way leads to surprising, often unintended results, challenging the common portrayal of inspiration as a form of guiding agent; instead, seeking to stimulate ideas through processes that go beyond the mind, as a means of creating original work. See Fell \cite[p.~21]{fell2022structure} and Magnusson and McLean \cite[p.~262]{magnussonmclean2018} for further exposition upon the value of harnessing processes that go beyond the imagination of the composer. We hold the belief that such a compositional attitude resonates particularly well with the aims of the VQH project, which similarly aims to unveil new musical territories by employing processes divergent from conventional musical practice.

Zen allows the user to map musical and sonic parameters across a three-dimensional canvas. Separate streams can be moved independently around this virtual space, with the parameters of sonic events being determined by their current position in time and space.
\color{black}

\subsubsection{Exporting data to Zen}
\label{subsubsec:zen_data}

The transmission of control data to Zen is done via HTTP. One can serve a data file or folder onto a simple HTTP server, so that Zen's web interface can poll data from, so long as they are connected to the same network. This is how the piece \textit{Dependent Origination} was technically realized. In summary, Zen can access data that is made available hosted locally or remotely through HTTP requests. The data is saved in a JSON file format, allowing Zen to parse the data by key. This process was improved for the piece \textit{Hexagonal Chambers} by hosting a purpose-built API in the cloud, allowing each performer to connect via WIFI, rather than a local network. This is discussed in more detail in the following section (\ref{subsec:hexagonal_chambers}).

\section{Composing with VQH}
\label{sec:composing}
In our capacity as composer-researchers, we have begun to explore the sonification of data generated using quantum algorithms through compositional processes.

Within the spectrum of quantum-computer assisted composition (QAC) we can highlight two distinct methods: either deferred, where musical forms and textures are carefully constructed outside of the act of performance, or in real-time, where improvisation is the prevalent process.

Acousmatic composition can be understood as a deferred compositional activity. Within this paradigm, the VQH can be used to manufacture, process, inter-connect, and inter-textualize a variety of sound objects in order to define the sonic and aesthetic landscape of a work. Conversely, the manipulation of synthesis parameters as a means of making the listener conscious of the interaction between artists during improvisation is a real-time activity, and one that positions the VQH as a musical instrument. Here, the interpreter learns to manipulate the QUBO coefficients to intentionally achieve musical variabilities. 

Quantum technologies have not yet developed a mature artistic language characteristic of their medium. However, the influence of quantum mechanics can be observed in the early history of electronic music; Dennis Gabor, Werner Meyer-Eppler, and others provided the theoretical basis for granular synthesis based on Gaussian particles \cite{gabor1947acoustical} \cite{roads2004microsound}. The use of granularity provides us with both the historical and aesthetic motivation with which to explore the application of the VQH within new musical terrains.

\paragraph{\textit{Rasgar, Saber}}
An initial example of an artistic output that used the VQH focused on exploring intermediary mappings (Sec. \ref{subsec:intermediary_maps}) to create sound objects used in an acousmatic composition. \textit{Rasgar, Saber} (Itabora\'i, 2023) is an acousmatic study that explores the concept of \textit{Quantum Itineraries of Sound}, as defined and thoroughly explored by \citet{itaborai2023towards}. In this work, though the main focus was the quantum representation of audio and the ways that audio signals can be translated between different media, there was an additional aesthetic concern being to establish a connection between early electronic music vocabulary and QAC. In this context, 4-qubit QUBOs were mapped into a granular synthesis generator driven by paper sounds, which were recorded and later composed as transitions between sections of the piece. Additionally, these objects were spatialized by VQH in four channels by using the strategy discussed in section \ref{subsubsec:spat_diffusion}. Both mappings were implemented in SuperCollider.

\subsection{Hexagonal Chambers}
\label{subsec:hexagonal_chambers}


Our most fruitful musical outcomes have been achieved by employing live coding software to shape and transform the data in an exploratory manner (Sec. \ref{subsec:live_coding}). We start with a direct mapping between data and sound, gradually loosening our approach as the composition unfolds and creative ideas emerge. 

In technical research, a frequent tension arises between the simultaneous needs to prove a hypothesis and allow for the emergence of more ambiguous aesthetic considerations. To remedy this, we often seek to locate the poetic amongst the empirical, emphasizing the need for academic rigour to be complemented by the ambiguity of metaphor, myth, and storytelling. By exploring literary narratives that evoke the technical processes we are researching, an encouraging feedback loop emerges: our research inspires our music, and in turn, our music inspires further research.

\textit{Hexagonal Chambers} (Thomas \& Itaborai, 2023) links the optimization process of variational quantum algorithms with a short story by Jorge Luis Borges; exploring the transition from chaos to order in all facets of musical language. The piece was performed using Zen, a live coding and data sonification system devised and built by Peter Thomas\cite{Zen} — discussed in the previous section — using data generated in real-time by Paulo Itaborai using the Variational Quantum Harmonizer.

In Jorge Luis Borges's short story "The Library Of Babel" \cite{borges1962ficciones}, the author explores the universe through the metaphor of a vast library. Comprised of a warren of hexagonal rooms, its catalog spans every possible permutation of language in a virtually infinite collection of incomprehensible texts. The library thereby contains all the knowledge ever to be written (and never to be written). Within this sea of chaos are the Vindications: books to absolve humanity of its sins, granting unity with God. And yet, due to the scale of the library, the search for these works — tomes of clarity amidst compendiums of noise, the light of meaning within a fog of meaninglessness — is considered to be futile.

In \textit{Hexagonal Chambers}, the performers act as librarians, with rooms represented by self-contained Zen algorithms, and books being separate datasets returned from the VQH. Leveraging the possibilities of live coding, the performers are able to execute and edit different algorithms on-the-fly, swapping out the underlying datasets and using them as the basis for further improvisation. This symbolizes the librarian's search through different rooms and texts. The composition's structure mirrors the optimization procedure found in the VQH algorithm, progressing from a state of meaningless noise to a stable truth.

\subsubsection{Generating Data}
\label{subsubsec:hc_data}

\begin{figure}[h]
    \centerline{\includegraphics[width=\linewidth]{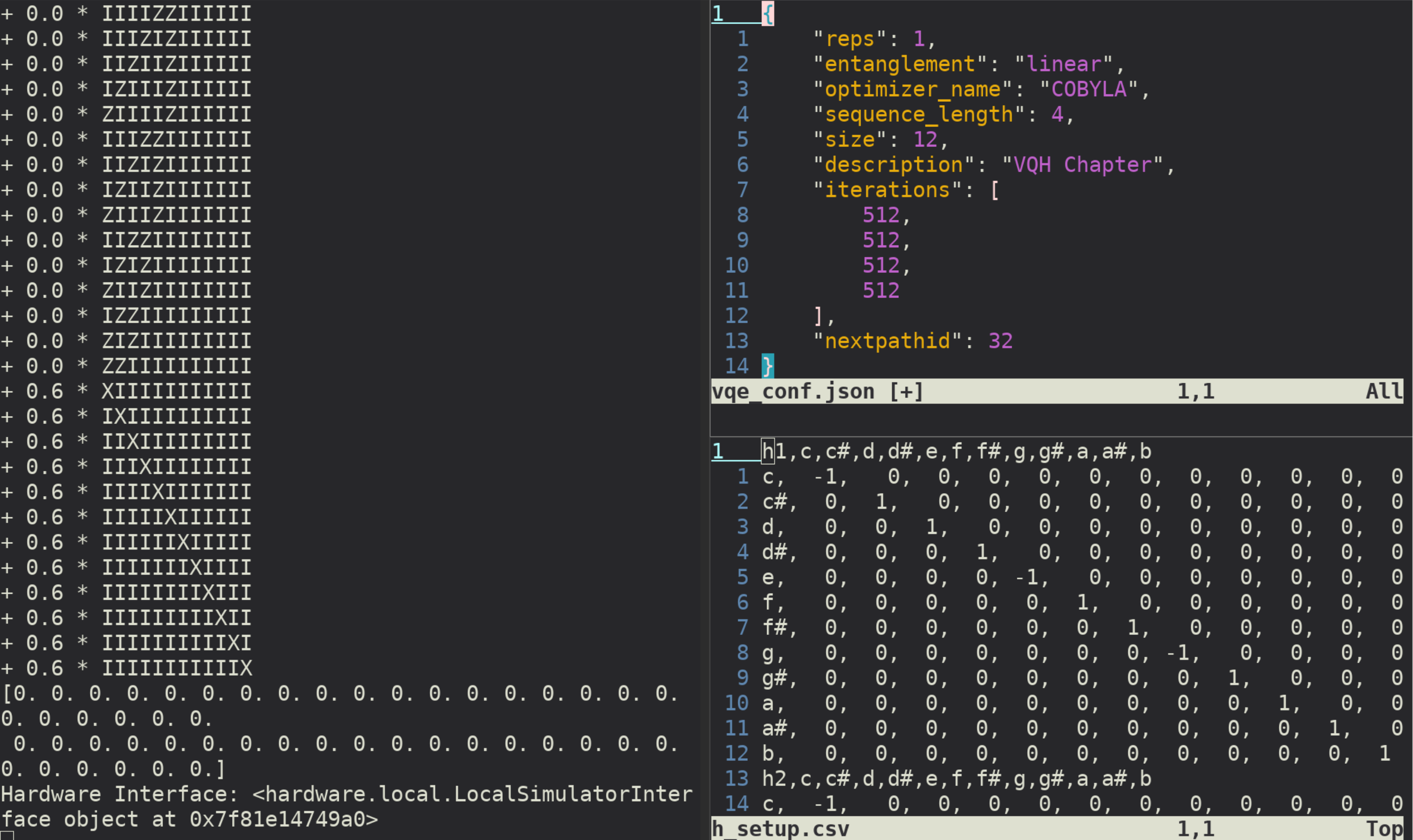}}
    \caption{VQH Interface for \textit{Hexagonal Chambers}}
    \label{fig:vqh_hc}
\end{figure}

The VQH is operated during the performance to generate new \textit{books}. Starting from simple linear QUBO matrices, and evolving in complexity to highly interconnected Ising systems with the inclusion of external Transverse Fields and non-linear coefficients (see sec. \ref{subsubsec:control_qubo}), a set of one to three new experiments was run for each section of the piece. Based on the current status of the piece, a new QUBO is designed and a classical optimizer selected. The choices are made with the intention to provoke musical gestures and variability. The parameters were controlled using a text-based approach, as the one illustrated in Fig. \ref{fig:vqh_hc}

As the VQH generates new data, they are uploaded to a purpose-built, remotely-hosted API. The VQH mapping function makes a HTTP \texttt{POST} call to an API endpoint.  The API accepts new datasets, or books, and stores them as single entries in the database. Zen was modified to be able to request new books via HTTP, as they are written.

Each new experiment generated a specifically designed dictionary containing all data mentioned in section \ref{subsec:son_data}, in JSON format.

As the mapping is triggered in VQH and the dictionary is sent to the remote server, Zen receives a notification, informing it of the arrival of a new \textit{book}.

\subsubsection{Sonifying Data}
\label{subsubsec:hc_son}

\begin{figure}[h]
    \centerline{\includegraphics[width=\linewidth]{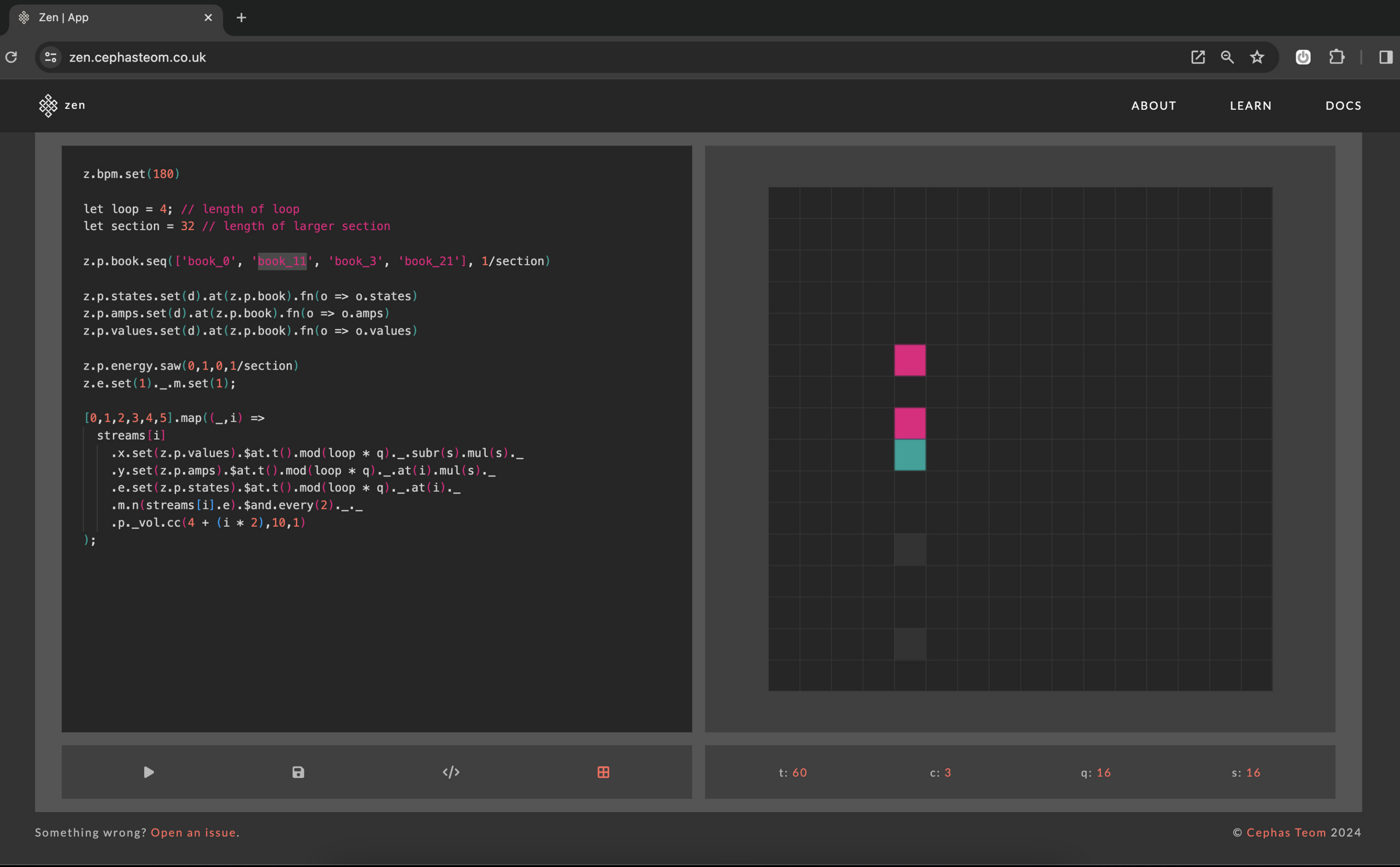}}
    \caption{Zen Interface for \textit{Hexagonal Chambers}}
    \label{fig:zen_hc}
\end{figure}

In our six-qubit system, we mapped the marginal coefficients to the xy-positions of the separate streams on the Zen canvas (Fig. \ref{fig:zen_hc}. Each stream was assigned to a separate instrument; which included samplers, and FM and granular synthesizers. Sonic events were triggered using the 6-bit binary string returned at each iteration, assigning one bit to each stream and triggering an event when the value was equal to 1. The resulting rhythms and spatial positioning were used as a basis for improvisation in real-time; experimenting with assigning different sonic parameters to each axis as the composition progressed.

\subsubsection{Syncing Zen and VQH}
\label{subsubsec:hc_sync_vqh}

\begin{figure}[h]
    \centerline{\includegraphics[width=\linewidth]{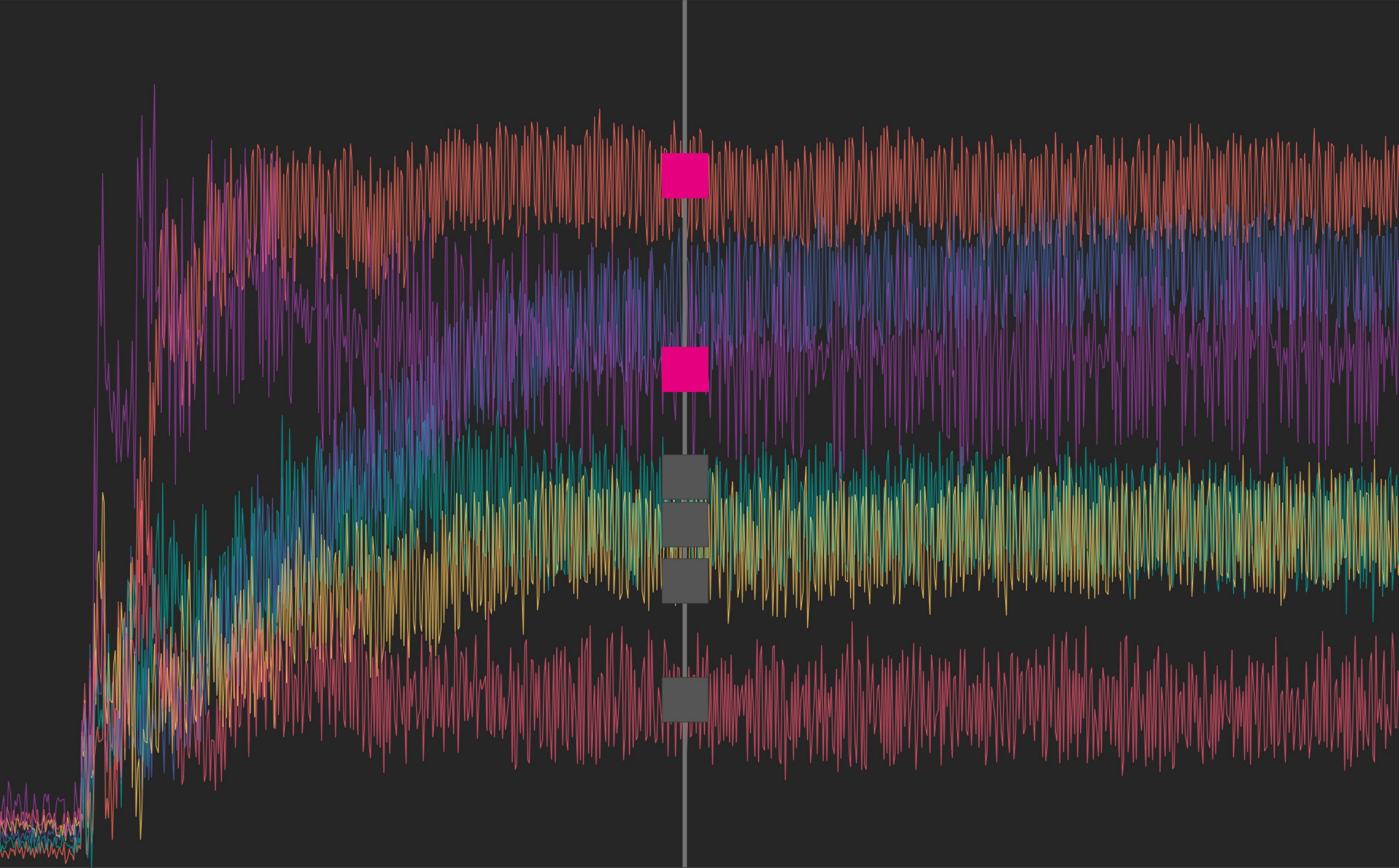}}
    \caption{Live visualization of the sonification data in Processing}
    \label{fig:vqh_zen}
\end{figure}

For the execution of the piece, it was decided that the audience should also visualize the data being sonified. To that end, a real-time visualization module, implemented in Processing \cite{fry2014processing}, was incorporated to the VQH system used in the performance. When a new \textit{book} was sent to the API, it also updated a local file used by processing to create a coloured plot of the marginal distribution data (Fig. \ref{fig:vqh_zen}). 

Furthermore, a MIDI connection was created between the machines running Zen and VQH. As a consequence, it was possible to send clock ticks from Zen's procedural step-generation as MIDI Control Change messages. This information was used to move a cursor in the visualization's plot, indicating which data points were driving the sound at each step. 

To send the time information as MIDI CC, which can span only 128 different values, the clock values had to be encoded into a stream of three consecutive MIDI messages, which could represent enough time steps in binary form to span the expected duration of the musical performance.

\section{Discussion}
\label{sec:discussion}

A novel framework for the sonification of Variational Quantum Algorithms was presented, and specific applications using a QUBO optimization problem and VQE minimization were shown. The implemented software interface, Variational Quantum Harmonizer, enabled artists to integrate quantum technologies in their compositional process, both in technical and aesthetic aspects, by using quantum principles as a musical form. These principles are used through quantum algorithms and methods that extend the traditional possibilities of procedural sound generation to the navigation of larger (Hilbert) spaces. The way a quantum computer processes information is profoundly different from the classical approach and requires its own musical language so that its manipulation can create meaningful sounds.

The VQH demonstrates preliminary pathways for the creation of audible visualization for Quantum Algorithms, by focusing on mapping sampled distributions of quantum states into sound. This structure could also be employed by researchers to sonify complex physical systems in the natural sciences to construct auditory displays of a problem of interest, potentially facilitating experiment workflows and the detection of new properties. Further, there are physical systems that are hard to compute with classical systems, such as the study of real-time phenomena in complex systems. 

These physical problems can benefit both from quantum computational approaches for simulation and from sonification. The elaboration of complex sonification mappings for large systems often involves artistic practices and practitioners who engage in interdisciplinary research. A noticeable example is the work on data sonification / musification of Wigner functions and Rabi Oscillations, realised within a prestigious institution of photonic research  \cite{ISQCMC_yamada}.

Moreover, VQH can be used as a powerful tool for training and education, specially when introducing Variational Quantum Algorithms. Sound and music can provide intuition to a broader range of individuals, even when they lack a formal musical background. For example, an audience in a concert can easily sing simple intervals or scales just by intuition. Thus, a deeper insight on how  Variational Quantum Algorithms optimize can be achieved by \textit{listening} to the process of optimization. This can be interpreted as a perceptual approach to gain understanding on abstract processes an ideas.

As a blueprint of a sonification toolbox, the VQH has a modular implementation, which enables for flexible scalability for the sonification of real complex problems, for circuit optimization and integration in different quantum computing cloud services and hardware, as well as for careful sound design tailored to a specific musical or scientific application. 

The implementation of the VQH is Open Source, and the source code, together with audio examples and code snippets, can be found and downloaded online \cite{VQHGit}.

\subsection{Future work}
\label{subsec:future_work}

At the time of writing, the authors are extending the implementation of the VQH for further research on the sonification. The following research directions are being currently addressed:

\begin{itemize}
    \item \textbf{Generalized OSC Communication}: The current implementation of VQH strongly reliant on SuperCollider and Zen. To become more flexible, a generic mapping is being developed, which dumps the sonification data with formatted Open Sound Control messages, which could be grabbed by any OSC-enabled software.
    \item \textbf{Musical Intrument Design}: Improvements on the implementation will permit the embedding of this sonification toolset in more encolsed systems, such as a Digital and Augmented Musical Instruments.
    \item \textbf{New Sonification Protocols}: There is an ongoing investigation on alternative qubit-efficient encoding methods and possibilities for mapping notes and musical features into quantum states. Two examples are Amplitude Encoding and Pauli-correlation Encoding \cite{sciorilli2024largescale}. Another range of encodings can be found in Quantum Representations of Audio \cite{itaborai2023towards}.
    \item \textbf{Sonification of High Energy Physics}: The group has been investigating the problem of sonifying Lattice Gauge Theories using VQH, such as the simulation of lattice QED in 2+1-dimensions \cite{Clemente_2022}.
    \item \textbf{Generalized Signals}: Sonification methods create one-dimensional signal streams, which are transformed into audio. However, in principle, this signal could be interpreted or transformed into other types of signals, from symbolic musical data, to images, ultrasound, microwaves, brain wave signals, and more.
\end{itemize}

\section{Acknowledgements}
\label{sec:acknowledgements}

P.I's and K.J.'s works are funded by the European Union’s Horizon Europe Framework Programme (HORIZON) under the ERA Chair scheme with grant agreement no.\ 101087126 (QUEST ERA Chair). 

K.J. and A.C.\ are supported with funds from the Ministry of Science, Research and Culture of the State of Brandenburg within the Centre for Quantum Technologies and Applications (CQTA). 
\begin{center}
\centerline{
    \includegraphics[width=0.1\textwidth]{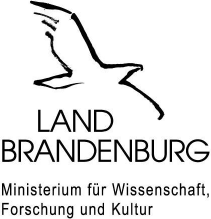}}
\end{center}

A. C. is supported in part by the Helmholtz Association
—“Innopool Project Variational Quantum Computer
Simulations (VQCS).”

T.S. is supported by the Einstein Research Unit -``Perspectives of a quantum digital transformation: Near-term quantum computational devices and quantum processors''

We thank Goethe-Institut for programming the piece \textit{Hexagonal Chambers} at the \textit{Wave Dysfunction} performance by Eduardo R. Miranda and QuTune Collective (including P.I, P.T. and M.Y.) during the 25th edition of the CTM Festival in Berlin, through their Studio Quantum Project. The event took place at Radialsystem, Berlin, 02 February 2024.

The VQH project started as a collaboration between the Interdiciplinary Centre for Computer Music Research at the University of Plymouth and the Center for Quantum Technologies and Applications at DESY, by initiative of professors Eduardo Reck Miranda and Karl Jansen.

\bibliographystyle{bst/ws-book-har}    
\bibliography{MAIN}      

\printindex

\end{document}